\documentclass{article}
\usepackage{graphicx}

\usepackage[margin=1in]{geometry}

\usepackage{setspace}
\usepackage{changepage}
\usepackage{enumitem}
\setlist{nosep}
\usepackage{caption}
\usepackage{natbib}
\usepackage{amsmath}
\usepackage{amssymb}
\usepackage{xcolor}
\usepackage{pifont}
\usepackage{rotating}
\usepackage{booktabs} 
\usepackage{url}

\usepackage{xcolor}
\usepackage{pifont}
\newcommand{\cmark}{\ding{51}}
\newcommand{\xmark}{\ding{55}}
\newcommand{\balpha}{\boldsymbol{\alpha}}
\newcommand{\bbeta}{\boldsymbol{\beta}}
\newcommand{\bgamma}{\boldsymbol{\gamma}}
\newcommand{\bepsilon}{\boldsymbol{\epsilon}}

\newcommand{\btheta}{\boldsymbol{\theta}}
\newcommand{\bvartheta}{\boldsymbol{\vartheta}}
\newcommand{\bphi}{\boldsymbol{\phi}}

\title{The Whittle likelihood for mixed models with application to groundwater level time series}
\author{Jakub J. Pypkowski$^{1,*}$, Adam M. Sykulski$^1$, James S. Martin$^1$, Ben P. Marchant$^2$}
\date{2 October 2026}

\begin{document}

\maketitle
\abstract{Understanding the processes that influence groundwater levels is crucial for forecasting and responding to hazards such as groundwater droughts. Mixed models, which combine a fixed mean, expressed using independent predictors, with autocorrelated random errors, are used for inference, forecasting and filling in missing values in groundwater level time series. Estimating parameters of mixed models using maximum likelihood has high computational complexity. For large datasets, this leads to restrictive simplifying assumptions such as fixing certain free parameters in practical implementations. In this paper, we propose a method to jointly estimate all parameters of mixed models using the Whittle likelihood, a frequency-domain quasi-likelihood. Our method is robust to missing and non-Gaussian data and can handle much larger data sizes. We demonstrate the utility of our method both in a simulation study and with real-world data, comparing against maximum likelihood and an alternative two-stage approach that estimates fixed and random effect parameters separately.}

\section{Introduction}
\let\thefootnote\relax\footnote{$^1$ Imperial College London, Department of Mathematics, London, SW7 2AZ, United Kingdom}
\let\thefootnote\relax\footnote{$^2$ British Geological Survey, Keyworth, NG12 5GG, Nottingham, United Kingdom}
\let\thefootnote\relax\footnote{$^*$ Corresponding author; email: jakub.pypkowski22@imperial.ac.uk}
Groundwater levels (GWLs) are recorded by measuring the water level in observation boreholes. A time series of GWL measurements from one borehole, called a groundwater hydrograph, reflects changes in the amount of water stored in a water-bearing layer of rock material called an aquifer. In particular, hydrographs can be used to detect and study groundwater droughts, a natural hazard defined as a period of substantially reduced availability of groundwater. Groundwater droughts may lead to costly restrictions of water supply for households, agriculture, and industry, as well as cause environmental damage \citep[][and references therein]{MARCHANT2018}. A successful response to groundwater droughts requires reliable forecasting, accurate warning systems, and sound water management plans \citep{vanLanen2016hydrology}. These require a good understanding of the processes driving droughts and necessitates models for GWLs.\par
Various assessment tools require that GWL data are collected at regular time steps and are complete [see e.g. \citealp{BloomfieldSGI2013}, \citealp{elhaj2025geotemporal}]. In reality, such standards are rarely achieved, and missing values must be filled in [\citealp{MARCHANT2018}, \citealp[]{Vu2021LSTM}]. Approaches to this vary from simple heuristics such as adopting a local mean, median, or linear interpolation between nearest observed values [\citealp[]{elhaj2025geotemporal}, \citealp[]{peterson2018interpolation}], to complex, physics-based flow models [e.g. \citealp[]{MACKAY2014}, \citealp[]{ZHOU2011}], which can also be used for inference and forecasting. The former approaches may not be scientifically justified in certain cases, while the latter models may be too complex to estimate when there are insufficient data, particularly when some predictor information is missing \citep{marchant2022temporal}. Recent developments in neural networks gave rise to Long Short-Term Memory models, which can fill in missing GWL values using few inputs, but do not provide insights into processes driving changes to GWLs [e.g. \citealp[]{Gholizadeh2023}, \citealp[]{Vu2021LSTM}]. \par
To bridge this gap, statistical modelling can be used. \cite{peterson2018interpolation} proposed a method which consists in estimating a nonlinear transfer function noise model \citep[TFN,][]{peterson2014TFN}, fitting an exponential covariance model to the variogram of the residuals, and using Kriging to interpolate. The TFN approach combines two ways of modelling GWLs: using independent meteorological variables, precipitation and potential evapotranspiration, as predictors, and modelling infiltration dynamics through a soil-moisture partitioning model \citep[for details see][]{peterson2014TFN}. Obtaining estimates of the model parameters involves solving an ordinary differential equation, and optimising an objective function given as a transformed sum of weighted squared innovations \citep{vonAsmuth2005modeling}. \par
An alternative approach using a mixed model was developed by \cite{MARCHANT2018} and \cite{marchant2022temporal}. They modelled GWLs as a sum of a fixed term and an autocorrelated random term and, similarly to \cite{peterson2018interpolation}, used Kriging to fill in missing values. The fixed term was expressed as a function of independent predictors including trends, seasonality, and meteorological variables (via impulse response functions), and then combined with random errors generated by a Gaussian process with an exponential covariance function. The model parameters were estimated using the Gaussian maximum likelihood. Both components of the model, fixed and random, were fitted simultaneously, unlike in \cite{peterson2018interpolation}, who estimated the fixed and random effects parameters in successive steps. The mixed model approach requires relatively few inputs and is simpler than physics-based models. \cite{MARCHANT2018} used the mixed model to group hydrographs across the Chalk aquifer in the UK by the main factors driving their variability and \cite{marchant2022temporal} confirmed that the mixed model performs well at filling in missing values. Because the mixed model incorporates a model for random errors, it allows for quantifying uncertainty. Through this, \cite{marchant2022temporal} also used it to detect outliers, and \cite{MARCHANT2018} proposed that it can be used for forecasting the risk of droughts.\par 
We highlight the distinction between the mixed models considered here, and mixed-effects models, where in the latter some coefficients of a (generalised) linear regression are treated as random and referred to as random effects. Linear mixed-effects models are expressed as $\mathbf{x}=M\bbeta+Z\mathbf{b}+\bepsilon$, where $M$ and $Z$ are known design matrices, $\bbeta$ is a vector of fixed but unknown coefficients called fixed effects, $\mathbf{b}$ is a vector of random effects, and $\bepsilon$ is a vector of independent error terms. We refer a reader interested in recent developments in scalable estimation of such mixed-effects models to \cite{Bellio2025}, \cite{Ghandwani2025}, \cite{Ghosh2022} for frequentist methods, and to \cite{Goplerud2025}, \cite{Papaspiliopoulos2020}, \cite{LoyaHrushikesh2025Asvi} for Bayesian methods.\par
There are two limitations to the state-of-the-art GWL mixed models we seek to address in this paper. The first is the assumption of Gaussian errors, where models with a more general distribution of the error term may provide a better fit to GWL data \citep{MARCHANT2016}. The second is the computational cost of maximum likelihood estimation, which increases steeply with the time series length $n$. This is due to the covariance matrix inversion in the Gaussian likelihood, which is $O(n^3)$ in general, with modern algorithms reducing it to approx. $O(n^{2.37})$ \citep[see e.g][]{alman2024matrixInv}. In the idealised case of a complete, regularly-sampled, and stationary time series, the covariance matrix becomes a Toeplitz matrix, inverting which has computational complexity of $O(n^2)$ \citep{trench1964toeplitz}. In practice, data are available at weekly, daily, or even 15-minute intervals and go as far back as the mid-19th century \citep{MARCHANT2018}. Maximum likelihood can therefore only be run on sub-sampled segments of such datasets. \par
To overcome these limitations, we propose a likelihood approximation method drawing from the Whittle likelihood \citep{whittle1953estimation}. The Whittle likelihood has become a common tool in estimating parameters of covariance models in second-order stationary time series. In this paper, we propose a Whittle-based method which simultaneously estimates all parameters of mixed models, can incorporate flexible forms of nonlinear means with impulse response functions, and can handle time series with missing data, all of which are important for GWL time series modelling and estimation in practice. This is done by integrating existing Whittle methods for regression-based models with recent developments for Whittle inference of stationary time series with missing data. Like the standard Whittle likelihood for stationary time series, our method does not require that random errors are Gaussian, only that they are fourth-order stationary. Similarly, our method retains the order $O(n\log n)$ of computational complexity of the standard Whittle likelihood. In a simulation study, we verify that our method can be successfully applied to data that exhibit characteristics typical for GWL measurements, including large volumes of missing data and non-Gaussianity.
\par
Several Whittle likelihood adaptations have been proposed for nonstationary processes [e.g. \citealp{dahlhaus2000likelihood}, \citealp[]{Li2021}, \citealp[]{Rosen2012}], but these works are focused on capturing nonstationarity in the covariance of the error term, for instance through local stationarity. They do not, however, consider a fixed nonstationary mean using independent variables. To this end, \cite{Ivanov2016Gauss} and \cite{Ivanov2020Levy} proposed a two-step procedure for models with a nonstationary mean driven by independent variables combined with Gaussian or L\'evy noise, respectively. In the first step of the procedure, the mean was estimated using least squares, and in the second step the Whittle likelihood was applied to the residuals. The advantage of our approach is that it simultaneously estimates the parameters of the fixed mean and the covariance of the error term, leading to reduced estimation error, as we shall show through a simulation study. Simultaneous estimation of all parameters using Whittle-based methods has been explored in several scenarios. \cite{Bertolacci2019} used it for models with an unknown fixed non-zero mean. \cite{koul2008testing} proposed a method for long-memory moving average processes, and \cite{Wang2015} and \cite{Huang2016} proposed a method for ARMA-type models, which was further adapted by \cite{WEI2022} and \cite{Baltagi2025} to estimation in panel data. \cite{goodwin2025} proposed a Bayesian inference method for models with long-range dependence of the error term. In all these approaches, the unobserved error term is modelled by subtracting the unknown mean, expressed as a function of free parameters, from the observed data. Such subtraction of the mean function is the key component of our framework, which we extend in this paper to handle time series with missing data as well as nonlinearity in the mean through impulse response functions.\par
Our method directly responds to challenges presented by GWL modelling, but its applicability extends beyond this area. First, filling in missing values is a common challenge in areas such as meteorology [e.g. \citealp[]{CoutinhoEluãRamos2023AoaC}, \citealp[]{rahman2024estimation}, \citealp[]{shabalala2019evaluation}], or for other types of time series data in hydrology, such as streamflow \citep{hiben2024estimation}, soil moisture \citep{MyeniL2022Davo}, and water quality monitoring \citep{ZhangYifan2022Hmdi}. Second, the method can be used with any time series where one is interested in modelling the mean, typically through regression, with autocorrelated error terms. Such problems are encountered in various areas of science, examples of which include pollution modelling \citep{yu2017pollution} and astrostatistics \citep{aigrain2023astrostat}.\par
The rest of this paper is structured as follows. We introduce the GWL data and present details of the mixed model in Section~\ref{sec:datamodel}. In Section~\ref{sec:estimation}, we formulate our framework for parameter estimation of a mixed model using the Whittle likelihood. We also outline two alternative estimation methods: Gaussian maximum likelihood and two-stage Whittle estimation. We perform a simulation study in Section~\ref{sec:sim}, in which we assess the performance of the presented estimation methods in various scenarios, which focus on challenges typical of GWL modelling including missing data and non-Gaussian observations. In Section~\ref{sec:app}, we model real-world GWL data using our Whittle-based framework. Lastly, we present our concluding remarks in Section~\ref{sec:conclusions}.

\section{Data and the mixed model}\label{sec:datamodel}
\subsection{Data}\label{sec:data}

\begin{figure}[!htb]
 \centering     
     \includegraphics[width=0.46\linewidth]{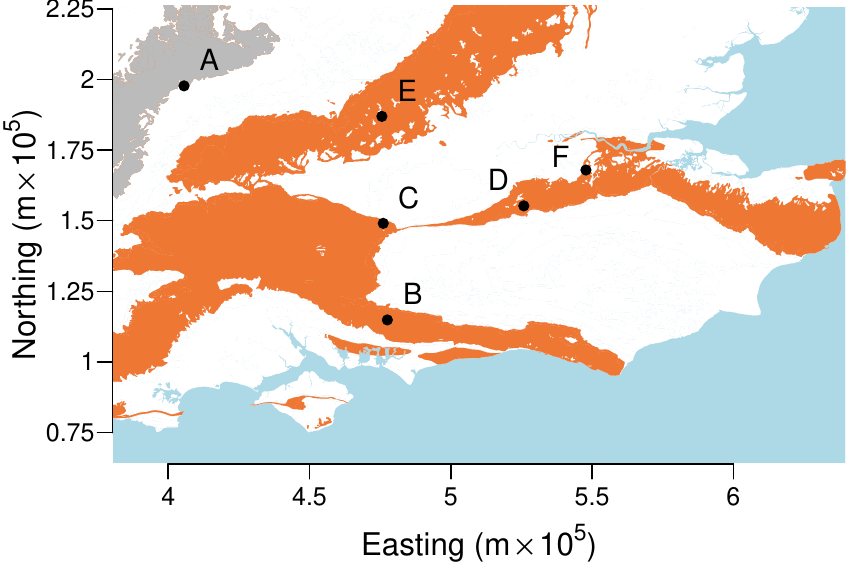}
 \caption{Locations of observation boreholes A--F across Southern England. Coloured areas show highly productive aquifers in which flow is virtually all through fractures and other discontinuities as per British Geological Survey 625K digital hydrogeological map (\protect\url{https://www.bgs.ac.uk/datasets/hydrogeology-625k/}). Great Oolite and Inferior Oolite groups are shown in grey. Other aquifers, the vast majority of which are part of the Chalk aquifer, are shown in orange. Contains British Geological Survey materials © UKRI 2025. Contains OS data © Crown copyright and database rights 2025. Contains public sector information licensed under the Open Government Licence v3.0.}
 \label{fig:map}
\end{figure}

\begin{figure}[!htb]
 \centering     
     \includegraphics[width=0.7\linewidth]{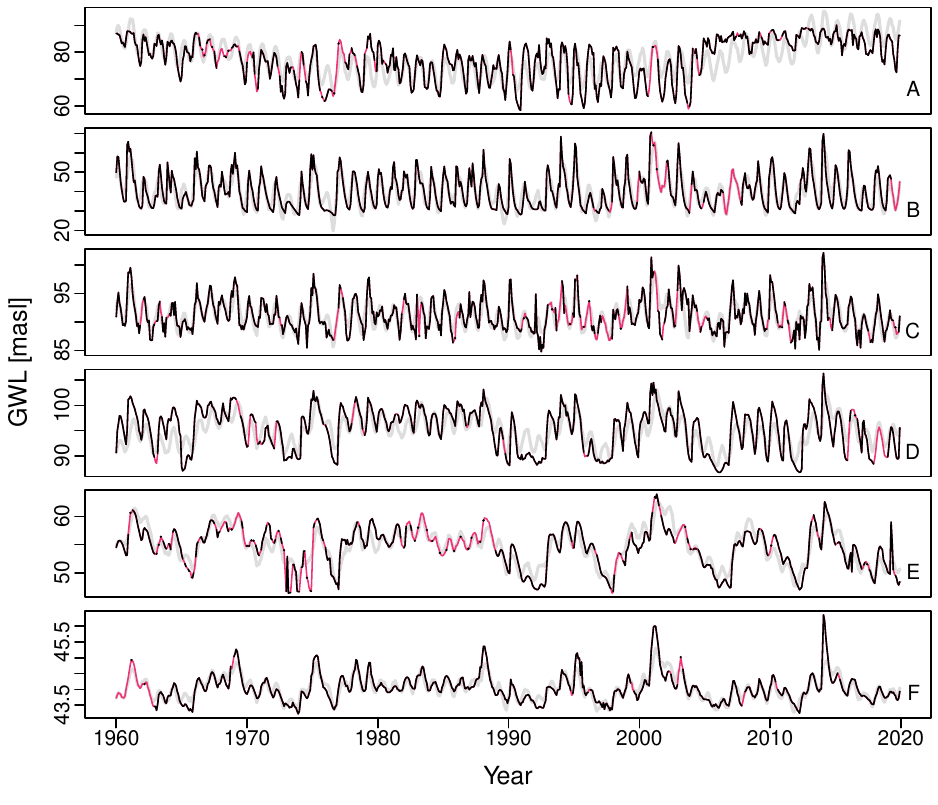}
 \caption{Recorded groundwater levels (black), missing value predictions (magenta) and estimated fixed-term mean (grey) for boreholes A--F (magenta and grey lines obtained using our Whittle-based method of Section~\ref{sec:estimation}). Black lines represent periods of no missing records, and black dots represent isolated observations. Contains data from the Environment Agency's Hydrology Data Explorer © Crown copyright 2025. Contains public sector information licensed under the Open Government Licence v3.0.}
 \label{fig:GWL_ts}
\end{figure}

We consider monthly GWL time series obtained from six observation boreholes across Southern England, which we identify with letters A--F. The time series span a period of 720 months from January 1960 to December 2019, and were obtained under the Open Government Licence v3.0 from the Environment Agency's Hydrology Data Explorer (\url{https://environment.data.gov.uk/hydrology/landing}). Figures~\ref{fig:map}--\ref{fig:GWL_ts} illustrate locations of the boreholes and corresponding time series, respectively. All time series exhibit yearly seasonality, which in the case of boreholes B and C dominates the overall variability of GWLs. In combination with longer-term trends (over a decade), seasonality is also a strong feature of the GWLs in borehole A. In boreholes D--F, seasonality gives way to medium-term features (up to 10 years). These patterns of variation within the hydrographs are broadly consistent with those observed by \cite{MARCHANT2018}, who performed a cluster analysis of standardised hydrographs in and around the Chalk aquifer. The Chalk aquifer is the major aquifer of southern and eastern England and includes the majority of the highly productive aquifers shown in Figure~\ref{fig:map}. One cluster contained boreholes from the Chilterns (in the vicinity of borehole E) and the eastern end of the North Downs (boreholes D and F). Consistent with Figure~\ref{fig:GWL_ts}, this cluster was characterised by relatively smooth hydrographs caused by deposits overlying the Chalk, which result in significant groundwater storage or delays to recharge. The cluster covering the South Downs (borehole B) was characterised by relatively rough hydrographs reflecting the highly fractured and faulted nature of this part of the aquifer. Borehole C, at the western end of the North Downs, corresponds to a cluster characterised by hydrographs with moderate smoothness. Borehole A lies in the Great Oolite aquifer, a Jurassic limestone aquifer. This aquifer is characterised by a combination of low storage ability (storativity) and high ability to transfer water (transmissivity) \citep{Neumann2003}. This makes GWLs in this aquifer highly responsive to meteorological factors and leads to large seasonal variation. \par 
The six time series are characterised by different missingness patterns. The fractions of data missing vary from 4.9\% for borehole D to 22.1\% for borehole E. The longest consecutive period of missing observations is 15 months, which occurred in the years 2006--2007 in borehole B. On the other extreme, there are 126 instances of records missing for only one month across the six time series. In previous studies of GWL time series \citep[e.g.][]{marchant2022temporal} the data were treated as missing completely at random (MCAR; meaning that whether a data point is observed or missing does not depend on any other variables). This assumption is reasonable, as the monitoring boreholes which are of interest are well-maintained and accessibility does not depend on groundwater levels. Missingness is primarily attributed to unexpected events independent of the data generating process, such as instrument failure, staff shortages, or external factors such as the foot-and-mouth disease outbreak which prevented manual data collection for many boreholes in 2001. If time series are collected from less well-maintained boreholes, the data may be missing at random (MAR; meaning that whether a data point is observed or missing depends on observed variables). For example, it is possible that instrument failure would become more likely over time, as the equipment deteriorates. This, however, is not suspected for boreholes A–F in our study. When the data are missing not at random (MNAR; meaning that whether a data point is observed or missing depends on other unobserved variables), the estimation methods considered in this paper may not be appropriate for estimating parameters of mixed models from GWL data. This may be the case when an overflow of groundwater leads to a local flooding, preventing manual data collection, or when the GWLs are outside of the range of the measuring equipment. These are cases of data being censored. Boreholes which suffer from such problems are not part of large-scale monitoring exercises, for which the mixed model presented in Section~\ref{sec:model} was designed.\par
To model GWLs, we use monthly rainfall data from \cite{rain_data} for locations of boreholes A--F from February 1950 to December 2019. We deseasonalise the rainfall data by removing a corresponding month's mean (across all years) from each observation.

\subsection{Mixed model}\label{sec:model}
In a mixed model, a response variable at a given time point is expressed as a sum of a {\em fixed} term, which describes the time-varying mean of the process as a possibly nonlinear function $\phi$ of a parameter vector $\bvartheta$, and a {\em random} term (or error) $\epsilon$, which is typically specified through its marginal distribution and covariance structure, the latter being controlled by the parameter vector $\balpha$. When applied to a regularly spaced time series, the model is given as
\begin{equation}\label{eq:model}
 \mathbf{x} = \bphi(\bvartheta) + \bepsilon, 
\end{equation}
where $\mathbf{x} = (x_1, \ldots, x_n)^T$ is the response vector, $\bphi(\bvartheta)=(\phi_1(\bvartheta), \ldots, \phi_n(\bvartheta))^T$ is a vector representing the time-varying mean as a function of $\bvartheta$, and $\bepsilon= (\epsilon_1, \dots, \epsilon_n)^T$ is a vector of random errors at times $1,\ldots,n$.\par
In this paper, we focus on the estimation of the mixed model developed by \cite{MARCHANT2018} and \cite{marchant2022temporal}. In their approach, the fixed term is assumed to be a linear function of terms including trend, seasonality, and a response to meteorological variables, and is written as $\bphi(\bvartheta) = M_{\bgamma}\bbeta$, where $M_{\bgamma}$ is an $n\times m$ design matrix containing $n$ values of $m$ covariates, $\bbeta$ is a vector of $m$ linear regression parameters, and $\bvartheta = \left(\bbeta^T, \bgamma^T\right)^T$. The fixed term is linear in $\bbeta$, however columns of the design matrix $M_{\bgamma}$ corresponding to the response to meteorological variables depend nonlinearly on the parameter vector $\bgamma$ through an impulse response function (IRF), as we shall explain later in this section. More specifically, the design matrix $M_{\bgamma}$ is built using the following covariates:
\begin{enumerate}
 \item An \textbf{intercept} encoded by a column of 1's in the design matrix $M_{\bgamma}$.
 \item A long-term \textbf{linear trend} encoded in $M_{\bgamma}$ as $(1/n, 2/n, \dots, 1)^T$.
 \item \textbf{Seasonal variation} encoded by two columns of $M_{\bgamma}$: one containing entries given by $\sin\left( 2\pi t/12\right)$ and the other by $\cos\left( 2\pi t/12\right)$, where $t$ is time (in months).
 \item \textbf{Temporary trends} encoded by several columns of $M_{\bgamma}$, each containing a cyclic spline basis function \citep[][p.202--204]{wood2017generalized}. Each basis function covers a different period and captures a deviation from the linear trend during a given period. Knots of the basis functions are spaced evenly. When the basis functions are used, the intercept becomes redundant and is excluded from the model.
 \item \textbf{Response to precipitation} encoded through an impulse response function \citep[IRF,][]{vonAsmuth2002IRF}. By using the IRFs, we account for a lagged response to precipitation, which is expected in GWLs. At the same time, we keep the number of model parameters low relative to an approach where one directly includes precipitation at multiple lags in the design matrix. IRFs aggregate the precipitation data, while assigning weights to precipitation at different lags. The weights reflect the importance of precipitation at a given lag for the GWL. The corresponding column $(m_1, \dots, m_n)^T$ of $M$ is given by 
 \begin{equation}\label{eq:IRF}
 m_t = \sum_{\tau=0}^{n_{IRF}-1} \text{IRF}(\tau)p(t-\tau),
 \end{equation}
 where $n_{IRF}-1$ is the maximum lag considered, and $p(t-\tau)$ is precipitation at time $t-\tau$. The weights assigned to an observation at lag $\tau$ are equal to the value of the chosen IRF function given by
 \begin{equation}
 \text{IRF}(\tau; A, s, a) = A\frac{a^{s}(\tau+1)^{s-1}\exp[-a(\tau+1)]}{\Gamma(s)}.
 \end{equation}
 \cite{marchant2022temporal} normalised the IRFs, by fixing $A$ so that the sum $\sum_{\tau=0}^{n_{IRF}-1} \text{IRF}(\tau; A, a, s) = 1$. We follow the same approach and as a result rewrite the entries $m_t$ given in~\eqref{eq:IRF} as 
 \begin{equation}\label{eq:IRFdata}
 m_t=m(t;s,a) = \frac{\sum_{\tau=0}^{n_{IRF}-1} (\tau+1)^{s-1}\exp[-a(\tau+1)] p(t-\tau)}{\sum_{\kappa=0}^{n_{IRF}-1} (\kappa+1)^{s-1}\exp[-a(\kappa+1)] }.
 \end{equation}
 The GWL depends on precipitation nonlinearly through the parameters $\bgamma=(s,a)^T$. We set $n_{IRF}=120$.
\end{enumerate}\par
We model the error term $\bepsilon= (\epsilon_1, \dots, \epsilon_n)^T$ as a realisation of a zero-mean continuous time process $\epsilon(t;\balpha)$ at integer times $(\epsilon(1,\balpha), \dots, \epsilon(n,\balpha))$. We denote the covariance of errors $\epsilon_t$, $\epsilon_{t+\tau}$, for any $t$, as $c_{\balpha} (\tau) = c_{\balpha} (\epsilon_t, \epsilon_{t+\tau})$. We consider a nested-nugget Mat\'ern covariance given as 
\begin{equation}\label{eq:matern}
 c_{\balpha} (\tau) = \begin{cases}
 c_0+c_1& \text{if } \tau=0,\\
 \frac{c_1}{2^{\nu-1}\Gamma(\nu)}\left(\frac{2\sqrt{\nu}\tau}{\lambda_m}\right)^\nu K_\nu\left(\frac{2\sqrt{\nu}\tau}{\lambda_m}\right) &\text{if } \tau >0,
 \end{cases}
\end{equation}
where $K_{\nu}(\cdot)$ is the modified Bessel function of the second kind of order $\nu$, and $\balpha=(c_0, c_1, \lambda_m, \nu)^T$ is a vector of parameters called: nugget, partial sill variance, length-scale, and smoothness, respectively. For some boreholes, we further consider a covariance function given as a product of Mat\'ern and periodic covariance functions (which we shall refer to as Mat\'ern-periodic), that is \eqref{eq:matern} multiplied by a periodic kernel $\exp\left(-2{\sin^2(\pi \tau/p)}/{\lambda_p^2}\right)$ if $\tau>0$, where $p$ is the period and $\lambda_p$ is the length-scale of the periodic kernel. This allows for random periodic effects to be captured if present in the data.\par

\cite{MARCHANT2018} and \cite{marchant2022temporal} assumed that $\epsilon(t,\balpha)$ is a Gaussian process in order to perform Gaussian maximum likelihood estimation. Our Whittle likelihood for mixed models (see Section~\ref{sec:ourWhittle}) allows us to relax this assumption: we do not specify a marginal distribution of the random term, instead we only require that it is fourth-order stationary. \cite{MARCHANT2018} and \cite{marchant2022temporal} pre-set the smoothness parameter of the Mat\'ern covariance to $\nu=1/2$ to reduce the number of parameters and speed up computation. This yields a simpler, exponential covariance, where $c_{\balpha}(\tau)=c_1\exp(-\tau/\lambda_m)$ if $\tau>0$. In this paper, we shall use our new methodology to estimate all four parameters of the Mat\'ern covariance, and for borehole A we shall estimate five parameters of the Mat\'ern-periodic covariance with a fixed period.

\section{The Whittle likelihood for mixed models}\label{sec:estimation}
In this section, we formulate our frequency-domain framework for estimating parameters of the mixed model. We first introduce the standard Whittle likelihood for second-order stationary processes \citep{whittle1953estimation}, which is commonly used for the estimation of covariance parameters of random processes with zero mean. We then introduce adaptations that allow for the estimation of parameters of the fixed term of the mixed model. We integrate two recent developments in the Whittle literature into our framework: modulation \citep{guillaumin2017modulated}, which allows for estimation in datasets with missing values, and debiasing \citep{sykulski2019debiased}, which improves accuracy of the estimation. Following the introduction of our method in Section~\ref{sec:ourWhittle}, we present two alternative estimation methods in Section~\ref{sec:alternatives}, including the Gaussian maximum likelihood. Once the parameters of the mixed model are estimated, they can be used for inference, filling in missing values, and forecasting. We perform the latter two using the simple Kriging predictor, for which we treat the estimated fixed term as a known mean [see \citealp[p.26--27]{van2019theory} for details].

\subsection{Method formulation}\label{sec:ourWhittle}
Let $Y(t; \balpha)$ be a zero-mean continuous-time Gaussian process with stationary covariance given by some covariance function $c_{\balpha} (\tau)$, and let $\mathbf{y} = (y(1, \balpha), \dots, y(n, \balpha)) = (y_1, \dots, y_n)$ be a finite, discrete-time realisation of the process at integer times. The spectral density function (sdf) of $Y(t; \balpha)$ for frequency $\omega_j$, when it exists, is given by
\begin{equation}
 f_{\balpha}(\omega_j) = \int_{-\infty}^\infty c_{\balpha}(\tau)\exp(-i\omega_j \tau)\text{d}\tau.
\end{equation}
The periodogram, an estimate of the sdf obtained using $\mathbf{y}$, is
\begin{equation}\label{eq:periodogramOG}
 I_j = \frac{1}{n}\left|\sum_{t=1}^n y_t \exp(-i\omega_jt)\right|^2,
\end{equation}
where $\omega_j \in \frac{2\pi}{n}(-\lceil n/2\rceil+1, \ldots, -1, 0, 1, \ldots, \lfloor n/2\rfloor)$. The log-likelihood for a Gaussian process can be written as 
\begin{equation}\label{eq:ML}
 l(\balpha) = \text{constant} - \frac{1}{2}\log|C(\balpha)| - \frac{1}{2}\mathbf{y}^TC(\balpha)^{-1}\mathbf{y}, 
\end{equation}
where $C(\balpha)$ is the covariance matrix of the discrete-time realisation $\mathbf{y}$.
The Whittle likelihood \citep{whittle1953estimation} approximates the log-likelihood in the frequency domain up to fixed constants, and is given in its discretised form by 
\begin{equation}\label{eq:whittleOG}
 l_W(\balpha) = -\sum_{j=1}^n\left[\log f_{\balpha}(\omega_j)+\frac{I_j}{f_{\balpha}(\omega_j)}\right].
\end{equation}
An argument $\hat \balpha$ which maximises $l_W(\balpha)$ is called a Whittle estimate. The computation of the Whittle likelihood is of order $O(n \:\log n)$ using the Fast Fourier Transform (FFT), which is lower than the computation of the Gaussian log-likelihood of \eqref{eq:ML}, as the latter requires matrix inversion. Another useful feature of the Whittle likelihood is that it is often robust to non-Gaussianity as the parameter estimates are consistent and asymptotically normal for all fourth-order stationary processes, as long as fourth-order moments are finite and fourth-order cumulants are absolutely summable [\citealp[]{dahlhaus1988small}, \citealp[]{Guillaumin2022DebiasedSpatial}, \citealp{sykulski2019debiased}]. This encompasses many stochastic processes that are not Gaussian.\par
We propose the following modification of the Whittle likelihood to adapt it for parameter estimation in a mixed model. Since the unobserved mixed model error process $\epsilon(t; \balpha)$ satisfies the conditions imposed on $Y(t;\balpha)$, we compute the periodogram of $\epsilon(t; \balpha)$ from the mixed model residuals $\mathbf{x} - \bphi(\bvartheta)$. The periodogram is therefore a function of the unknown fixed term parameter vector $\bvartheta$ given as
\begin{equation}
 I_j(\bvartheta) = \frac{1}{n}\left|\sum_{t=1}^n\left[x_t - \phi_t(\bvartheta) \right]\exp(-i\omega_jt)\right|^2.
\end{equation}
We replace $I_j$ in \eqref{eq:whittleOG} by $I_j(\bvartheta)$, and yield the Whittle likelihood expressed jointly as a function of $\balpha$ and $\bvartheta$:
\begin{equation}\label{eq:whittleSubtracted}
 l_W(\balpha, \bvartheta) = -\sum_{j=1}^n\left[\log f_{\balpha}(\omega_j)+\frac{I_j(\bvartheta)}{f_{\balpha}(\omega_j)}\right].
\end{equation}
We call arguments $\hat \balpha$ and $\hat \bvartheta$ which maximise $l_W(\balpha, \bvartheta)$ Whittle estimates of $\balpha$ and $\bvartheta$.\par
In addition to being computationally efficient through the use of the FFT, the Whittle likelihood has desirable asymptotic properties for fourth-order stationary residuals. In finite samples, however, it may exhibit poor performance, in particular leading to a large estimation bias [see e.g. \citealp{dahlhaus1988small}, \citealp{sykulski2019debiased}]. This results from spectral blurring, an issue common in spectral analysis \citep{walden1993}, caused by the truncation of stochastic processes to finite-length samples. \cite{sykulski2019debiased} proposed a modification of the Whittle likelihood called debiasing, and showed that it can substantially reduce bias in finite samples. It consists in replacing the sdf in \eqref{eq:whittleOG} with the expected periodogram given by
\begin{equation}
 \bar{f}_{\balpha}(\omega_j) = \mathbb{E}(I_j)=2 \operatorname{Re}\left[\sum_{\tau=0}^{n-1}\left( 
 1-\frac{\tau}{n} \right) c_{\balpha}(\tau) \exp(-i\omega_j\tau)\right]-c_{\balpha}(0),
\end{equation}
which can be computed via an FFT. We apply the same modification to the proposed Whittle likelihood for mixed models in \eqref{eq:whittleSubtracted}. \par
To be fit for our purpose, the Whittle likelihood must also allow for missingness in the data. To ensure this, we further refine our method by combining it with a framework proposed by \cite{guillaumin2017modulated} on modulated time series. Let $\mathbf{g} = (g_1, \dots, g_n)$ be a modulating sequence such that
\begin{equation}\label{eq:modseq}
 g_t = \begin{cases} 
 1 &\text{if the value of } x_t \text{ is observed}, \\
 0 &\text{otherwise}.
 \end{cases}
\end{equation}
Then the modulated time series is given as $\Tilde{\mathbf{x}}$ $ = (\Tilde{x}_1, \dots, \Tilde{x}_n) = (x_1g_1, \dots, x_ng_n)$, meaning it takes the value of $x_t$ if it is observed, or $0$ if it is not. 
Analogous to \cite{guillaumin2017modulated}, we introduce the following modifications to our Whittle likelihood for mixed models. We first replace the periodogram of residuals with a periodogram of modulated residuals
\begin{equation}
 \tilde{I}_j(\bvartheta) = \frac{1}{n}\left|\sum_{t=1}^n g_t\left [x_t - \phi_t(\bvartheta) \right]\exp(-i\omega_jt)\right|^2,
\end{equation}
and then obtain its expectation as 
\begin{equation}\label{eq:expModulated}
 \tilde{f}_{\balpha}(\omega_j)=\mathbb{E}[\tilde{I}_j(\bvartheta)]=2 \operatorname{Re}\left[\sum_{\tau=0}^{n-1}\tilde{c}_{\balpha}(\tau) \exp(-i\omega_j\tau)\right]-\tilde{c}_{\balpha}(0),
\end{equation}
where $\Tilde{c}_{\balpha}(\tau)$ is the expectation of the biased sample autocovariance of the modulated process, given by
\begin{equation}
 \tilde{c}_{\balpha}(\tau) = c_{\balpha}(\tau)\frac{1}{n} \sum_{t=0}^{n-\tau-1} g_tg_{t+\tau}.
\end{equation}
Then, our final debiased Whittle likelihood for mixed models which accounts for missing data is given by the following formula:
\begin{equation}\label{eq:whittleFinal}
 l_W(\balpha,\bvartheta) = -\sum_{j=1}^n\left[\log\tilde{f}_{\balpha}(\omega_j)+\frac{\tilde{I}_j(\bvartheta)}{\tilde{f}_{\balpha}(\omega_j)}\right].
\end{equation}
This form of the Whittle likelihood retains the original computational complexity of $O(n \: \log n)$ thus offering a substantial computational speed-up relative to maximum likelihood. The proposed Whittle likelihood is reliable under MCAR or MAR for sufficiently large sample sizes. The case of MAR requires that the variables on which the missingness probability depends are included as model predictors. This reliability under both MCAR and MAR follows from the fact that the Whittle likelihood approximates the likelihood function, and maximum likelihood is known to produce asymptotically unbiased estimates under these missingness regimes \citep[p.134]{missingData_txtbook}. Inference for data that are MNAR remains an active area of research, but neither the maximum likelihood nor our Whittle method should be expected to perform well under this regime. We refer a reader to \cite{missingData_txtbook} for a thorough overview of missingness regimes.\par
\cite{Guillaumin2022DebiasedSpatial} show that under Gaussianity, the confidence intervals can be asymptotically approximated by the inverse Hessian. We provide code in our GitHub repository that allows users to easily approximate such confidence intervals for large sample sizes. In a general case (without requiring Gaussianity), the covariance matrix of the Whittle estimates can be asymptotically found as 
\begin{equation}
    \operatorname{Cov}\left\{ \left( \widehat{\balpha}^T, \widehat{\bvartheta}^T\right)^T\right\}  = \mathcal{H}^{-1}\left\{ \left( \balpha^T, \bvartheta^T\right)^T\right\} \operatorname{Cov}\{\nabla l_W(\balpha, \bvartheta)\}    \mathcal{H}^{-1}\left\{ \left( \balpha^T, \bvartheta^T\right)^T\right\} ,
\end{equation}
where $\mathcal{H}^{-1}\left\{ \left( \balpha^T, \bvartheta^T\right)^T\right\}$ is the inverse Hessian, and $\operatorname{Cov}\{\nabla l_W(\balpha, \bvartheta)\}$ is the covariance matrix of the Whittle likelihood score function. We refer the reader to \cite{Guillaumin2022DebiasedSpatial} for further details on how the covariance can be approximated efficiently in such cases.\par
The proposed Whittle likelihood for mixed models can be further refined to handle nonstationarity in the random term. One approach is to consider the error process as locally stationary, that is to assume that the covariance parameters vary over time, but that they do so slowly enough that one may consider them constant over some short periods of time. In the frequency domain, such an approach was taken by \cite{Sykulski2016Drifters} who used a rolling window of observations to estimate the time-varying parameters using the Whittle likelihood. Applied to mixed models, this approach has the disadvantage that parameters of the fixed term would also have to be estimated over the same rolling windows, even if they do not vary over time themselves. To overcome this issue, an alternative approach is to use a generalised version of the \cite{guillaumin2017modulated} framework that we used to handle missing data in~\eqref{eq:modseq}. For example, we can address heteroskedasticity by assuming that random errors have constant variance, but are then rescaled by the modulating sequence $\mathbf{g}=(g_1, \dots, g_n),$ $g_t\in\mathbb{R}^{+}\cup\{0\}$, which can now take values other than 0 or 1, and is either assumed completely known, or known up to some functional form whose parameters are estimated alongside all other parameters of the model. Then, the Whittle likelihood for mixed models takes the exact same form as~\eqref{eq:whittleFinal} and can estimate all parameters across the whole time series at once. Changes in periodicity can also be incorporated by using a complex-valued modulation function (see \cite{guillaumin2017modulated} for details).

\subsection{Alternative estimation methods}\label{sec:alternatives}
\subsubsection{Gaussian maximum likelihood} \label{sec:exactGaussLik}
Maximum likelihood estimation is a common tool across various fields of statistics due to its estimates' desirable properties, such as consistency and efficiency. The method consists in maximising the likelihood function (or optimising its appropriate transform). For the mixed model, the Gaussian likelihood function is derived from the properties of $\epsilon (t; \balpha)$ assumed to be a Gaussian process with covariance function $c_{\balpha}(\tau)$. The Gaussian log-likelihood, which is the generalisation of \eqref{eq:ML} for a mixed model, is given by
\begin{equation}\label{eq:exact}
 l(\balpha, \bvartheta) = \text{constant} - \frac{1}{2}\log|C(\balpha)| - \frac{1}{2}(\mathbf{x} - \bphi(\bvartheta))^TC(\balpha)^{-1}(\mathbf{x} - \bphi(\bvartheta)),
\end{equation}
and can be maximised to estimate the parameters $\balpha$ and $\bvartheta$ simultaneously. \par 
As in general there is no closed-form solution to the optimisation of the likelihood function, it instead must be optimised numerically. \cite{marchant2022temporal} show, however, that for the GWL model specification (with $\bphi(\bvartheta)=M_{\bgamma}\bbeta$), conditional on $\balpha$ and $\bgamma$, the function is optimised when 
\begin{equation}\label{eq:beta}
 \bbeta = (M_{\bgamma}^T C(\balpha)^{-1}M_{\bgamma})^{-1}M_{\bgamma}^T C(\balpha)^{-1}\mathbf{x}.
\end{equation}
Due to this, we only need to numerically explore the space of parameters $\balpha$ and $\bgamma$, and then obtain $\hat \bbeta$ analytically given the estimates $\hat \balpha$ and $\hat \bgamma$. This speeds up the optimisation, but \cite{marchant2022temporal} note that the estimation may still be computationally expensive for large datasets due to the computational complexity of matrix inversion $\geq O(n^2)$. Maximum likelihood estimation accounts for missing data: for any unobserved entry $x_i$ of $\mathbf{x}$, one removes the $i$-th row of the design matrix $M_{\bgamma}$, and the $i$-th row and column of $C(\balpha)$.\par

\subsubsection{Two-stage estimation with the Whittle likelihood}
We follow the approach by \cite{Ivanov2016Gauss}, who proposed a two-stage estimation procedure for models composed of a regression component and a Gaussian error. The approach combines least squares estimation for the regression parameters with Whittle estimation of the covariance parameters. \cite{Ivanov2016Gauss} formulated the method in terms of continuous-time processes. We present a simplified two-stage procedure for processes sampled at discrete times.\par
In the first stage, we obtain the estimate $\hat \bvartheta$ as an argument that minimises the least squares expression $(\mathbf{x}-\bphi(\bvartheta))^T(\mathbf{x}-\bphi(\bvartheta))$. For the GWL model specification (with $\bphi(\bvartheta)=M_{\bgamma}\bbeta$), conditional on $\bgamma$, the expression is minimised by $\bbeta = \hat{\bbeta}=(M_{\bgamma}^TM_{\bgamma})^{-1}M_{\bgamma}^T\mathbf{x}$, as per standard least-squares results. Using this property, one only needs to optimise numerically for $\bgamma$. We note that the least squares method coincides with maximum likelihood when the errors are i.i.d. Gaussian. \par
In the second step, we use model residuals $\hat \bepsilon = \mathbf{x} - \bphi(\widehat{\bvartheta})$ to obtain a Whittle estimate $\hat \balpha$ of the covariance parameters. We employ the debiasing and modulated process frameworks described in Section~\ref{sec:ourWhittle} to improve performance with missing data. We obtain the periodogram of the modulated residuals as
\begin{equation}
 \Tilde{I}_j = \frac{1}{n}\left|\sum_{t=1}^n g_t\hat{\epsilon}_t \exp(-i\omega_jt)\right|^2,
\end{equation}
and its expectation is analogous to \eqref{eq:expModulated}. We then optimise the Whittle likelihood given by \eqref{eq:whittleFinal}.\par
The two-stage estimation may significantly reduce the computational cost of parameter estimation, especially as the Whittle likelihood, rather than maximum likelihood, is used to estimate $\balpha$. Nevertheless, it comes with a theoretical disadvantage: by not estimating all parameters simultaneously one adds some simplifying assumptions that may reduce the accuracy of the estimation. In particular, the first stage of the method implicitly assumes the independence of error terms, which we explicitly assume to be correlated. This may reduce the accuracy of the estimation of the mean, particularly in areas with many missing observations. Our novel method overcomes these issues by estimating parameters jointly, as proposed in Section~\ref{sec:ourWhittle}, and we shall contrast these approaches in our simulation study in Section~\ref{sec:sim}.

\section{Simulation study}\label{sec:sim}
We conduct a simulation study to assess the performance of the three estimation methods presented in Section~\ref{sec:estimation}, including our novel Whittle likelihood for mixed models. We focus on aspects of the performance such as estimation and prediction errors. We consider three scenarios that reflect practical questions of interest in GWL modelling. In the first scenario, we consider the model presented in Section~\ref{sec:model}. Through this, we assess how well different estimation methods may perform on typical GWL data. In the second scenario, we fit the same model to data generated fully randomly, i.e. with the fixed term set to zero. In this way, we assess whether the estimation method can appropriately indicate variables that are not relevant predictors of GWLs. Lastly, we consider a model with non-Gaussian errors to assess how robust the estimation methods are to model misspecification. In all simulation study scenarios, as well as in Section~\ref{sec:app}, we optimise the log-likelihood, Whittle likelihood, and least-squares functions using the Broyden-Fletcher-Goldfarb-Shanno (BFGS) algorithm implemented using the \texttt{optim} function in \texttt{R}. Following the method of the \texttt{fminsearchbnd} function in MATLAB \citep{fminsearchbnd}, we refined the algorithm by wrapping the parameter space to implement lower and upper bounds on the parameters wherever necessary.
\subsection{Simulation study design}\label{sec:simDesign}
For each scenario, we consider four lengths of time series $n\in \{256,\: 512,\: 1024,\: 2048\}$, where 25\% of observations are missing completely at random. In all cases, we use the three estimation methods from Section~\ref{sec:estimation}, which we henceforth abbreviate to Whittle, maximum likelihood (ML), and 2-stage. Additionally, for the third scenario (with non-Gaussian errors), we use maximum likelihood with a correctly specified likelihood function for the non-Gaussian model from which we generate data. The main components of the data generating process in the simulation are visualised in Figure~\ref{fig:simulationExamples} with further details on each scenario as follows:
\begin{sidewaysfigure}[!htpb]
 \centering
 \includegraphics[width=\linewidth]{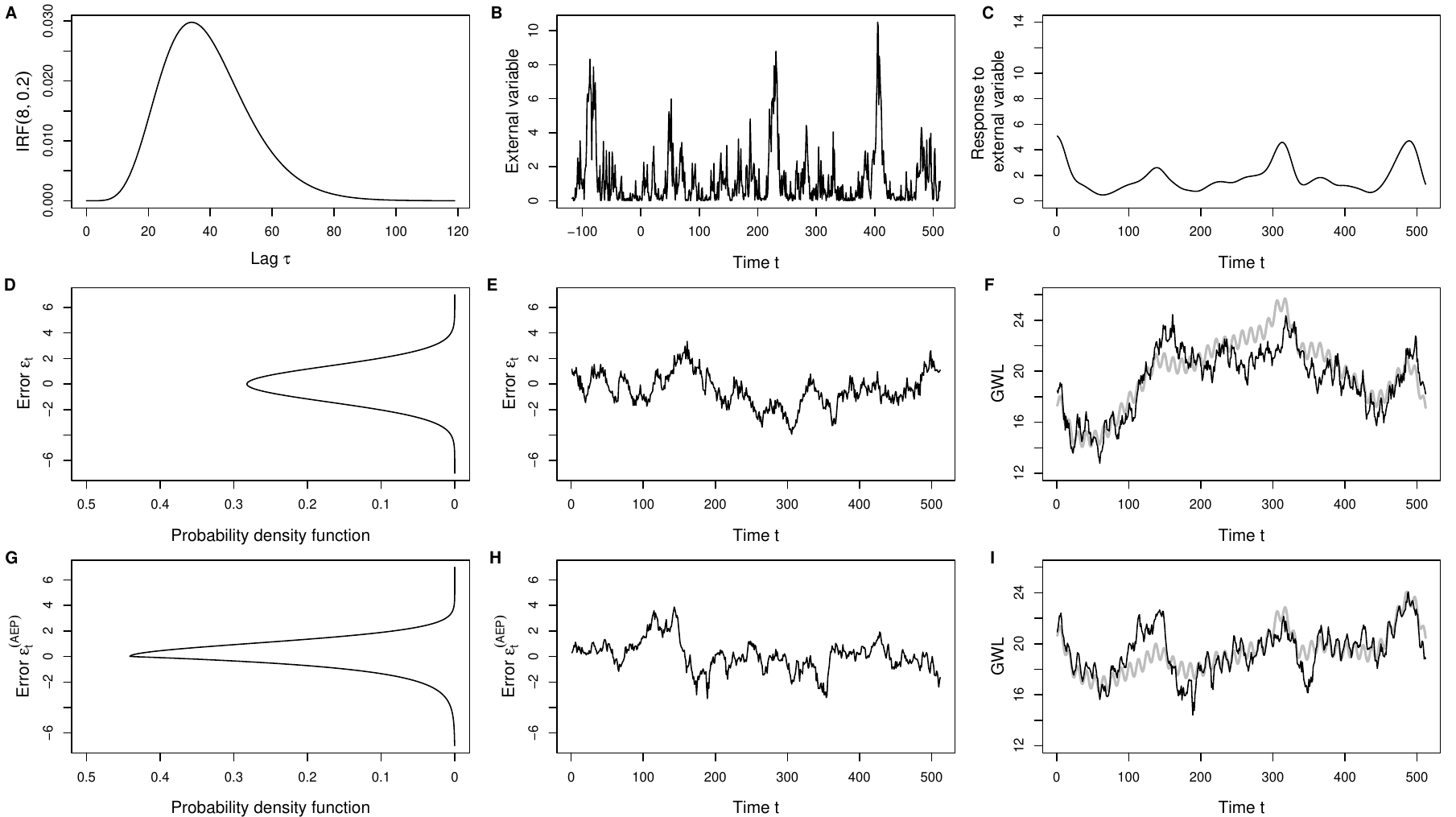}
 \caption{Visualisation of generation of time series in the simulation study. \\ Panel A: the IRF used in our study. Panel B: a realisation of the external variable. Panel C: the impulse response for the realisation presented in panel B.\\
 Panel D: a pdf of the Gaussian error $\epsilon_t$. Panel E: a realisation of the Gaussian error $\bepsilon$. Panel F: the fixed term (gray), including the response to the external variable realisation in panel C, and the observed GWL (black) with Gaussian error presented in panel E.\\
 Panels G--I: as in panels D--F, except for the model with AEP-distributed errors $\bepsilon^{(AEP)}$.}
 \label{fig:simulationExamples}
\end{sidewaysfigure}
\begin{enumerate}
 \item A \textbf{standard mixed model} is specified as the model presented in Section~\ref{sec:model}. Specifically, we incorporate a linear trend, seasonal variation, four cyclic splines, and a response to an external variable (e.g. rainfall). The parameter vector (in the order of these features) is $\bbeta = (3,\: 0.15,\: -0.6,\: 18,\: 10, \:18,\: 20, \:1.5)^T$. The external variable is generated by exponentiating a zero-mean Gaussian process with nested nugget Mat\'ern covariance with parameters $\balpha_{ext}=(0.1,\: 1,\: 15,\: 0.5)^T$. We set $n_{IRF}=120$ and generate $n+119$ values of the external variable. A realisation of the external variable is presented in panel B of Figure~\ref{fig:simulationExamples}. The parameters controlling the IRF are $\bgamma=(8,\:0.2)^T$; the function is plotted in panel A. In panel C, we present the GWL response to the realisation of the external variable. The error term is simulated as a Gaussian process with Mat\'ern covariance with parameters $\balpha=(0.05, \:2,\: 25,\: 1.5)^T$. The probability density function (pdf) of the error's marginal distribution is presented in panel D, and a realisation of the error time series is presented in panel E. Panel F combines all these components: the grey line represents the fixed term with the presented impulse response, and the black line presents a realisation of GWL time series with the presented realisation of errors. In total there are 14 parameters to be estimated across $\{\balpha, \bbeta, \bgamma\}$.

 \item A \textbf{fully random model} is specified by setting all entries of $\bbeta$ to 0. That is, such a model consists of random errors only and panel E in Figure~\ref{fig:simulationExamples} presents a realisation of this model. We fit a standard mixed model to the simulated data, i.e. we also estimate parameters $\bbeta$ and $\gamma$. For the estimation, we use the same design matrix $M_{\bgamma}$ and external variable as in the standard mixed model scenario. 
 
 \item A \textbf{mixed model with asymmetric exponential power distribution} \citep[AEP,][]{Zhu2009AEP} is specified by changing the marginal distribution of the error term. Such a model was used in the context of GWL modelling by \cite{MARCHANT2016} to account for potential skewness and heavy tails of the distribution of the error term. To impose a correlation structure onto the model, they used a Gaussian copula and a Mat\'ern covariance. We follow the same approach. The AEP distribution is parameterised by five parameters: a location parameter $\mu\in\mathbb{R}$, a scale parameter $\sigma>0$, a skewness parameter $\varsigma\in(0,1)$, and left and right tail parameters $p_1>0$, and $p_2>0$, respectively. For the purpose of this paper, we set $\mu=0$, but note that this does not lead to a zero-mean process, as the parameter $\mu$ is equal to the mode, not the mean of the process \citep{Zhu2009AEP}. As the dispersion of the distribution is controlled by the parameter $\sigma$, we fix the total variance of the underlying Gaussian process to $c_0+c_1=1$ (see~\eqref{eq:matern}). We simulate the AEP-distribution errors using the inversion method \cite[][p.51]{Copulas2006}, which we present step by step in Appendix~\ref{apx:AEP}.\par
 In this scenario, we use a simpler form of the fixed term of the mixed model to account for the greater number of parameters describing the distribution of the error term and still estimate 14 free parameters. Namely, we leave out the spline basis functions corresponding to variation in the long-term trend. Instead, we add an intercept term with parameter $\beta_1=16$. We split the unit variance of the underlying Gaussian process between the nugget $c_0=0.1$ and the partial sill $c_1=0.9$. We keep the remaining parameter values the same as in the standard mixed model scenario, and additionally set $\btheta=(\sigma, \varsigma, p_1, p_2)^T$ s.t. $\sigma = 1.4, \varsigma = 0.4$, $p_1=1$, $p_2=1.9$, so that the left side of the distribution has heavier tails, but, on average, only two negative errors are expected for every three positive ones. The pdf of the marginal distribution of the error is presented in panel G of Figure~\ref{fig:simulationExamples}, and a realisation of the error time series is presented in panel H. Panel I combines the realisation of the error with the fixed term. As in other scenarios, we fit the mixed model assuming Gaussian errors to data generated with AEP-distributed errors. In this way, we assess how robust the methods are to model misspecification. Additionally, we use a correct log-likelihood function which does account for the AEP-distributed errors, and which we present in Appendix~\ref{apx:AEP}. 

\end{enumerate}

In Table~\ref{tab:metrics}, we present the metrics we use to compare the performance of estimation methods in our simulation study. As the scenarios are all different, the appropriate set of metrics varies across the study, as depicted in the table and explained here in more detail. Where appropriate, we assess the accuracy of estimation of parameters $\balpha$ and $\bbeta$ using absolute proportional errors averaged for all parameters in each vector, in other words we use the $L^1$ norm of the estimation errors divided by the true values of parameters $||(\hat \cdot -\cdot)/\cdot||_1$ (rows 1 and 3 of Table~\ref{tab:metrics}). For the model with AEP-distributed errors, the parameters fitted with the assumption of Gaussianity cannot be compared meaningfully with the true values. Instead, we consider a metric of divergence between the true and estimated autocovariance functions $c_{\balpha}(\cdot)$, $c_{\hat{\balpha}}(\cdot)$ of the error term: $\text{Div}[ c_{\hat\balpha}(\cdot)] =\sum_{\tau=0}^{n-1} |c_{\hat\balpha}(\tau) - c_{\balpha}(\tau)|$ (row 2 of Table~\ref{tab:metrics}).
For the fully random scenario in which $\bbeta=\mathbf{0}$, relative estimation errors cannot be computed as the truth is zero. Instead, we consider unscaled errors $||(\hat \bbeta -\bbeta)||_1$ (row 4 of Table~\ref{tab:metrics}). For mixed models, we do not compare parameter estimates for $\bgamma$ directly, as they are not of immediate interest. Rather, we compare the divergence metrics between the estimated and true IRFs: $\text{Div}[IRF(s,a; \hat s, \hat a)] =\sum_{\tau=0}^{n_{IRF}-1} |IRF(\tau; \hat s, \hat a) - IRF(\tau; s, a)|$ (row 5 of Table~\ref{tab:metrics}). We compare the accuracy of filling in missing values using the root mean squared error (RMSE) for all missing observations $\mathbf{x}_{miss}$ (row 6 of Table~\ref{tab:metrics}), which are predicted using simple Kriging. In the scenario with AEP-distributed errors, simple Kriging is applied to model residuals converted to the Gaussian distribution (see Appendix~\ref{apx:AEP}). Then, using the inversion method, the predicted values are converted back to the AEP distribution and added to the estimated fixed term. For all metrics, lower values indicate better performance. 

\begin{table}[htb]

\caption{Performance metrics used to compare estimation methods in three simulation scenarios.}

\centering
\begin{tabular*}{\columnwidth}{@{\extracolsep\fill}lccc@{\extracolsep\fill}}
\toprule
& \multicolumn{3}{@{}c@{}}{Simulation scenario} \\
\cline{2-4}
 Metric & Standard Mixed & Fully Random & AEP--distributed error \\
 \midrule
$||(\hat{\balpha}-\balpha)/\balpha||_1$ & {\color{teal}\cmark} & {\color{teal}\cmark} & {\color{magenta}\xmark} \\
$\text{Div}[ c_{\hat\balpha}(\cdot)]$
& {\color{magenta}\xmark} & {\color{magenta}\xmark} & {\color{teal}\cmark} \\
$||(\hat{\bbeta}-\bbeta)/\bbeta||_1$ & {\color{teal}\cmark} & {\color{magenta}\xmark} & {\color{teal}\cmark} \\
 $||(\hat{\bbeta}-\bbeta)||_1$ & {\color{magenta}\xmark}
& {\color{teal}\cmark} & {\color{magenta}\xmark} \\
$\text{Div}[IRF(s,a; \hat s, \hat a)]$ & {\color{teal}\cmark} & {\color{magenta}\xmark} & {\color{teal}\cmark} \\
$\text{RMSE}(\hat{\mathbf{x}}_{miss})$ & {\color{teal}\cmark} & {\color{teal}\cmark} & {\color{teal}\cmark}\\
\hline
\end{tabular*}
\label{tab:metrics}
\end{table}

\begin{figure}[htb]
 \centering
 \includegraphics[width=0.84\linewidth]{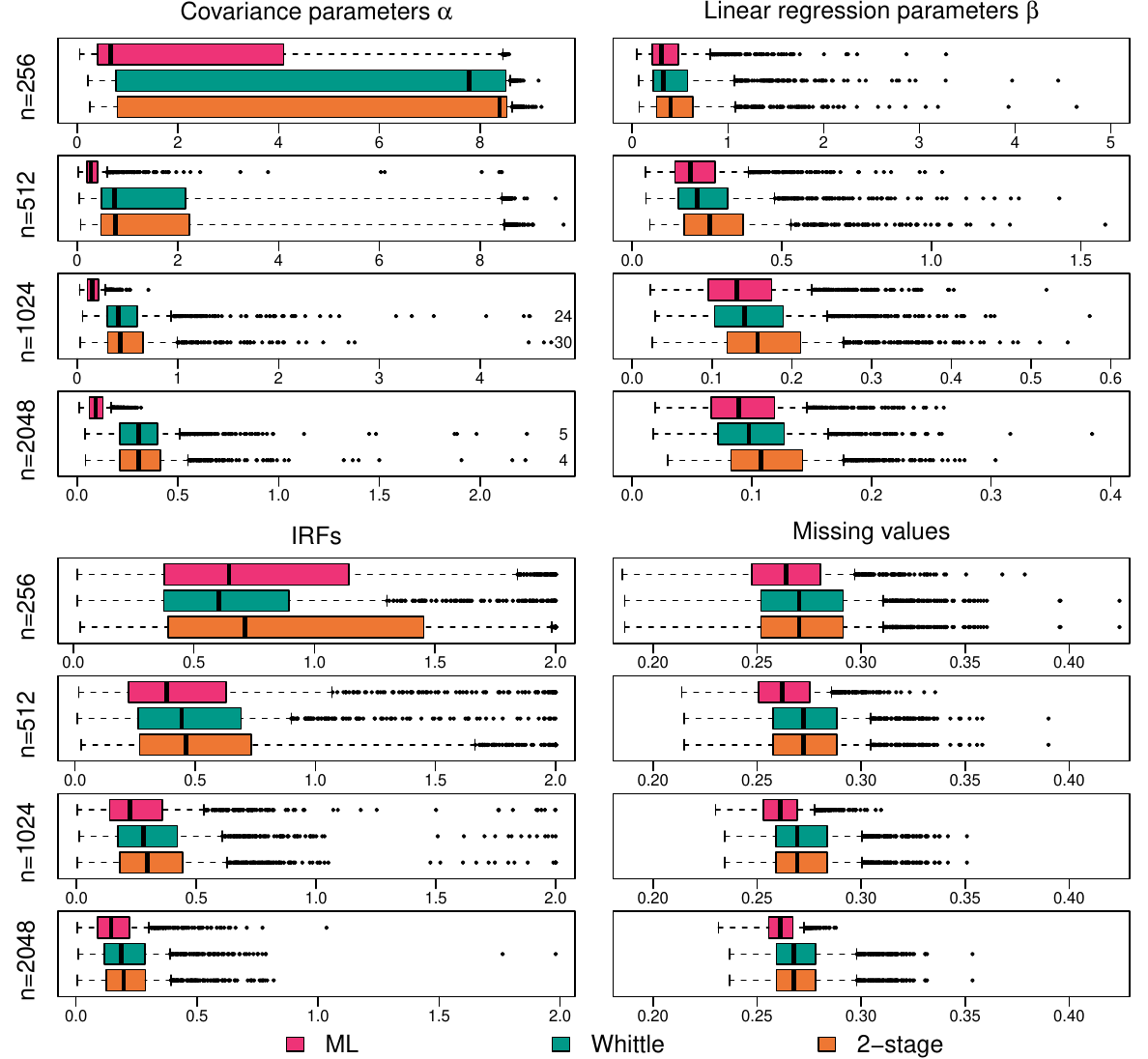}
 \caption{Simulation results for the standard mixed model scenario: boxplots of $L^1$ relative errors of covariance parameter estimates $\hat\balpha$ and linear regression parameter estimates $\hat\bbeta$, divergence between true and estimated IRFs, and RMSE of prediction of missing values.}
 \label{fig:sim_Gauss}
\end{figure}
\subsection{Simulation study results}\label{sec:sim_results}
\begin{figure}[t]
 \centering
 \includegraphics[width=0.84\linewidth]{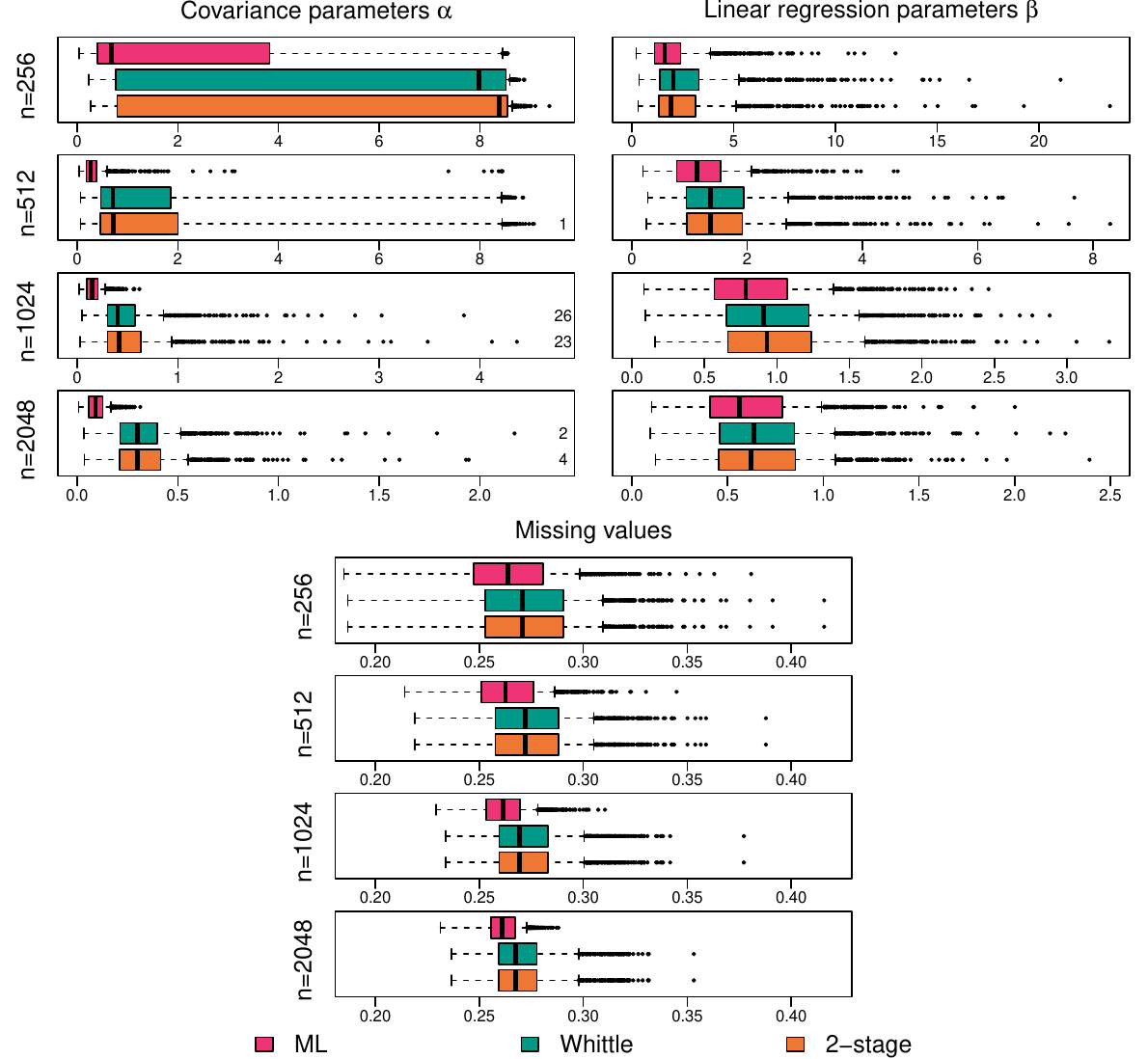}
 \caption{Simulation results for the fully random model scenario: as in Figure~\ref{fig:sim_Gauss} except with absolute (instead of relative) $L^1$ errors of covariance estimates $\hat{\bbeta}$ and without metrics for IRF divergence.}
 \label{fig:sim_Zero}
\end{figure}

\begin{figure}[t]
 \centering
 \includegraphics[width=0.84\linewidth]{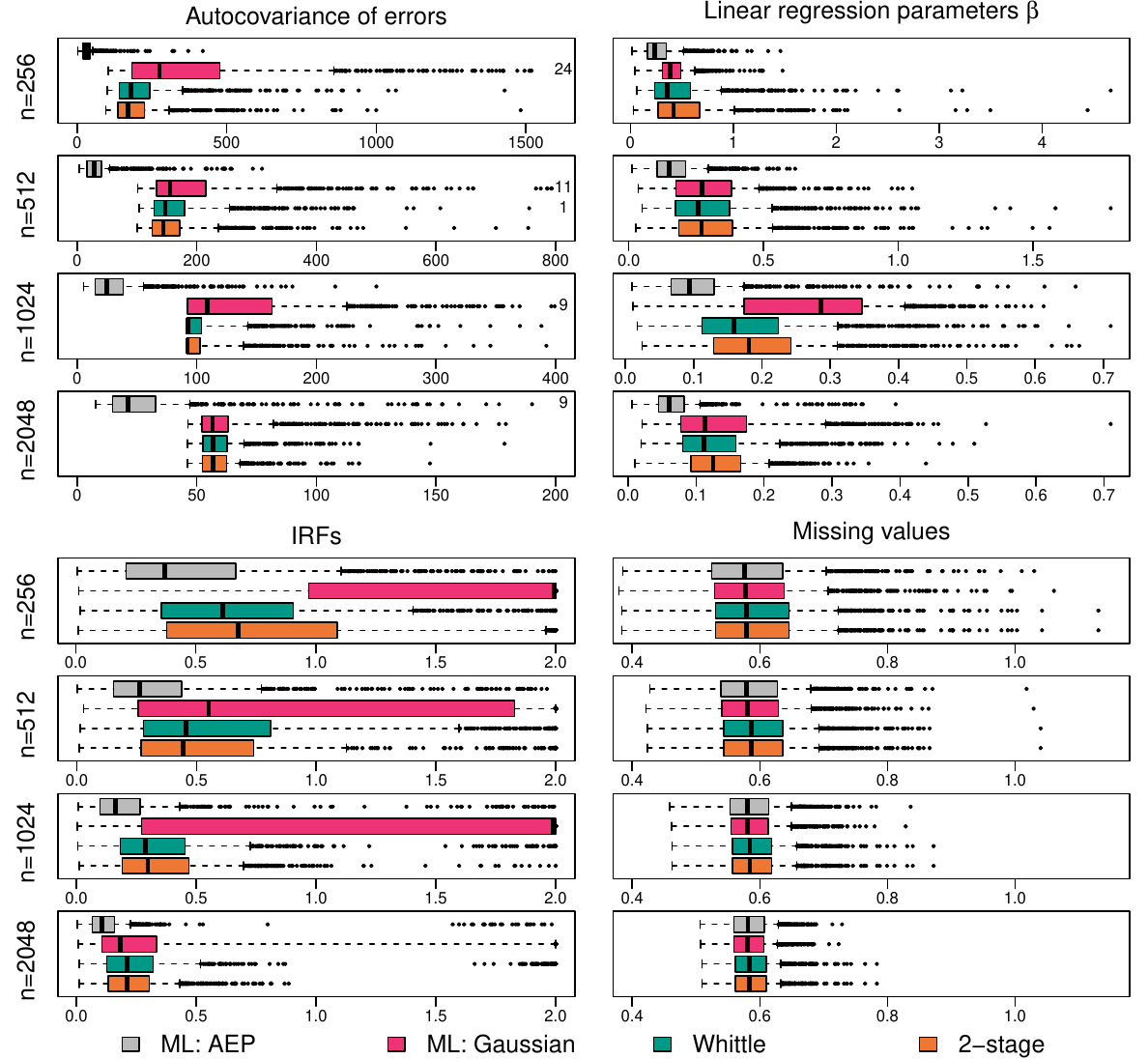}
 \caption{Simulation results for the AEP-distributed errors scenario: as in Figure~\ref{fig:sim_Gauss} except with divergence between true and estimated autocovariance of random errors instead of estimation errors of covariance parameters.}
 \label{fig:sim_AEP}
\end{figure}

Performance metrics from the simulation study are presented in Figures~\ref{fig:sim_Gauss}--\ref{fig:sim_AEP}, each figure showing metrics for one simulation study scenario. The results are presented as boxplots, where each datapoint represents a performance metric from one simulation run. The whiskers of boxplots represent the minimum and the 90th percentile of the recorded metric values. The highest 10\% of the values are represented as dots. In cases where most extreme metric values would have compromised the legibility of the other elements of the boxplots, we cut the highest values off and report the number of points that are not displayed. Below we discuss the results by comparing the performance of different estimation methods across the three scenarios. \par

As expected, the correctly specified \textbf{maximum likelihood}, i.e. with Gaussian errors in the scenarios with standard mixed model and fully random model, and with AEP-distributed errors in the remaining scenario, outperformed other methods across all metrics in nearly all cases. Few exceptions from this occurred only for cases with low numbers of observations, e.g. for the IRF divergence in the standard mixed model scenario (Figure~\ref{fig:sim_Gauss}). The advantage of maximum likelihood was generally greater for the metrics related to the random error than to the fixed term of the mixed model. The advantage was smallest for the RMSE of the missing value prediction. At the same time, the advantage was overall greater in the scenario with AEP-distributed errors. Here, missing values prediction was an exception, as all methods performed similarly well.\par
In the scenario with AEP-distributed errors (Figure~\ref{fig:sim_AEP}), maximum likelihood specified for Gaussian errors performed worse than Whittle and 2-stage. This disadvantage is visible in terms of metrics related to autocovariance of the random error, regression parameters $\bbeta$, as well as IRFs. This demonstrates that Whittle and 2-stage were more robust to non-Gaussianity than using the (misspecified) Gaussian maximum likelihood in this study. Despite this, the Gaussian maximum likelihood did not perform worse than Whittle and 2-stage when it came to missing value prediction, as also shown by \cite{MARCHANT2016}.\par
\textbf{Whittle} and \textbf{2-stage} performed similarly well at predicting missing values in each scenario. The advantage of Whittle over 2-stage is most visible in the standard mixed model scenario (Figure~\ref{fig:sim_Gauss}), which represents a typical case of GWL data. Here, Whittle slightly outperformed 2-stage at the estimation of covariance parameters $\balpha$, and more substantially of the regression parameters $\bbeta$. In the latter case, the performance metrics of Whittle were closer to maximum likelihood than to 2-stage. When it comes to divergence of true and estimated IRFs, the advantage of Whittle was larger for shorter time-series and diminished with increasing $n$. In the fully random model scenario (Figure~\ref{fig:sim_Zero}), there was a slight advantage of Whittle over 2-stage in terms of the estimation of covariance parameters $\balpha$. Unlike in the standard mixed model scenario, here both methods performed similarly well when regression parameters $\bbeta$ were concerned, meaning that the two methods had a similar ability to discern insignificant predictors of the fixed term of the mixed model. In the scenario with AEP-distributed errors (Figure~\ref{fig:sim_AEP}), the performance metrics for Whittle had better central percentiles (25th, 50th, 75th) than 2-stage when one considers the estimation of linear regression parameters $\bbeta$ or IRFs (except when $n=512$). At the same time, the tails of the distributions of the metrics were a bit heavier for Whittle. In terms of the remaining metrics, both methods resulted in a similar performance. For the shorter time series, Whittle exhibited a slight disadvantage in terms of divergence of true and estimated autocovariance of random errors.\par
The results of the study demonstrated that Whittle can be successfully applied to typical GWL data with missing values. It offered an improvement over 2-stage while being more computationally efficient than maximum likelihood. For AEP-distributed errors, the two Whittle-based methods were more robust to non-Gaussianity than using a misspecified maximum likelihood assuming Gaussian errors.

\section{Application}\label{sec:app}
We now fit the mixed model to the GWL time series presented in Section~\ref{sec:data}.
\begin{table}[t]
\centering
\caption{Comparison of average run times of the estimation using maximum likelihood and Whittle for various covariance models for boreholes A--F in \texttt{R} (4.3.2) on a desktop computer with a 12th Gen Intel(R) Core(TM) i5-1245U (1.60 GHz) processor.}
\begin{tabular}{llcc}
\hline
 & & \multicolumn{2}{@{}c@{}}{Boreholes}\\
\cline{3-4}
 Estimation method  &  Covariance model & A       & B--F     \\
\hline
ML      & Exponential     & 74.15s  & 82.57s  \\
ML      & Mat\'ern          & N/A & 645.22s \\
Whittle & Mat\'ern          & N/A & 53.35s  \\
ML      & Mat\'ern-periodic &  691.72s    & N/A     \\
Whittle & Mat\'ern-periodic &  54.35s & N/A    \\
\hline
\end{tabular}

\label{tab:runTimes}
\end{table}
We fit models with a Mat\'ern-periodic (borehole A; with a yearly period) and Mat\'ern (boreholes B--F) covariance using Gaussian maximum likelihood and Whittle, as well as a model with exponential covariance using the maximum likelihood. Using a simpler covariance function with Gaussian maximum likelihood reflects standard practice, and follows the modelling approach taken by \cite{MARCHANT2018} and \cite{marchant2022temporal} who simplified the covariance function to speed up estimation. For each borehole, we ran the optimiser 20 times with different initial values to ensure convergence to the global minimum, as local minima occur for all models and estimation methods---we encourage users of mixed models to take a similar approach independent of the estimation method they choose. In Table~\ref{tab:runTimes}, we present average run times. For full models (Mat\'ern-periodic or Mat\'ern), using Whittle reduces the average optimisation time by a factor of 12.2 relative to using maximum likelihood. This translates to a time reduction of estimation for one borehole (20 optimiser runs) from, on average, 3h 38min to only 18min. This shows that our Whittle-based method significantly outperforms maximum likelihood in terms of computational effort. In what follows, we compare the output of the full models estimated using Whittle against the simpler, standard-practice model with exponential covariance estimated using maximum likelihood. Such comparison is fairer in terms of required computational expense than comparing full models estimated using the two methods, but we note that maximum likelihood still required longer computation than Whittle despite fitting a simpler model. We find, as expected given the results of our simulation study, that maximum likelihood and Whittle lead to very similar mean (fixed term) estimates and missing value predictions. We present the estimates and predictions obtained using Whittle in Figure~\ref{fig:GWL_ts}. \par
\begin{figure}[!b]
 \centering
 \includegraphics[width=0.94\linewidth]{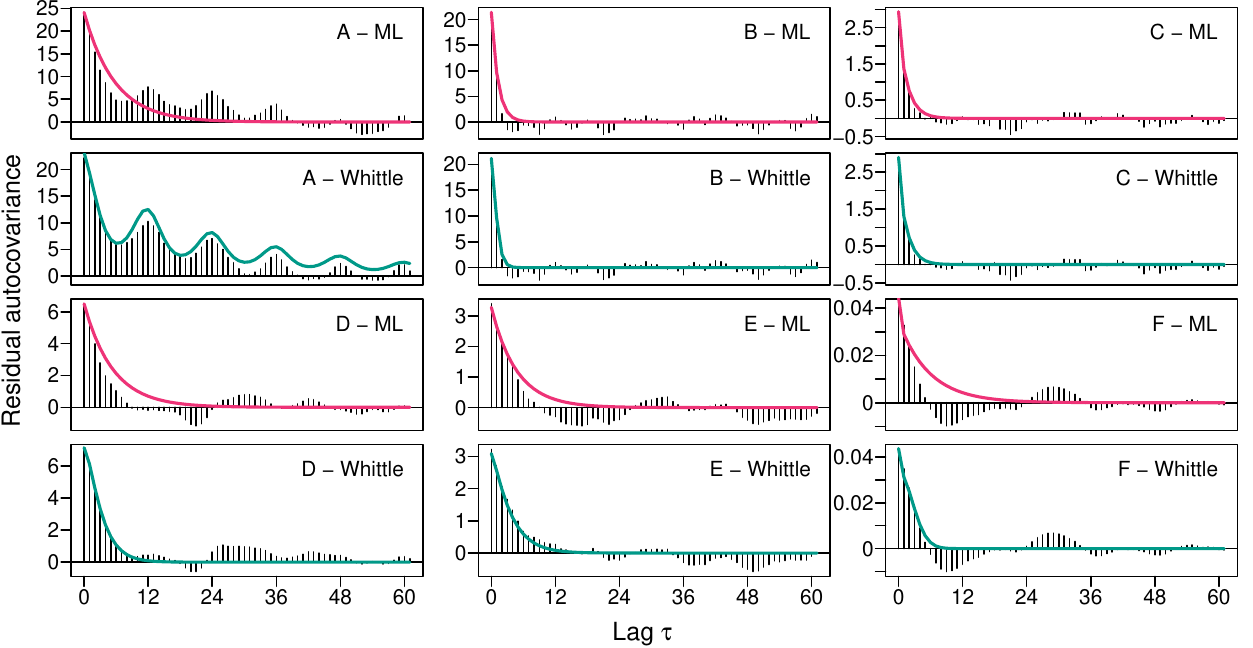}
 \caption{Empirical autocovariance of model residuals (bars) and covariance functions using parameter estimates obtained using maximum likelihood (magenta, exponential covariance) and Whittle (teal, Mat\'ern and Mat\'ern-periodic covariance) for boreholes A-F.}
 \label{fig:acv_model}
\end{figure}

\begin{figure}[!htb]
 \centering
 \includegraphics[width=0.94\linewidth]{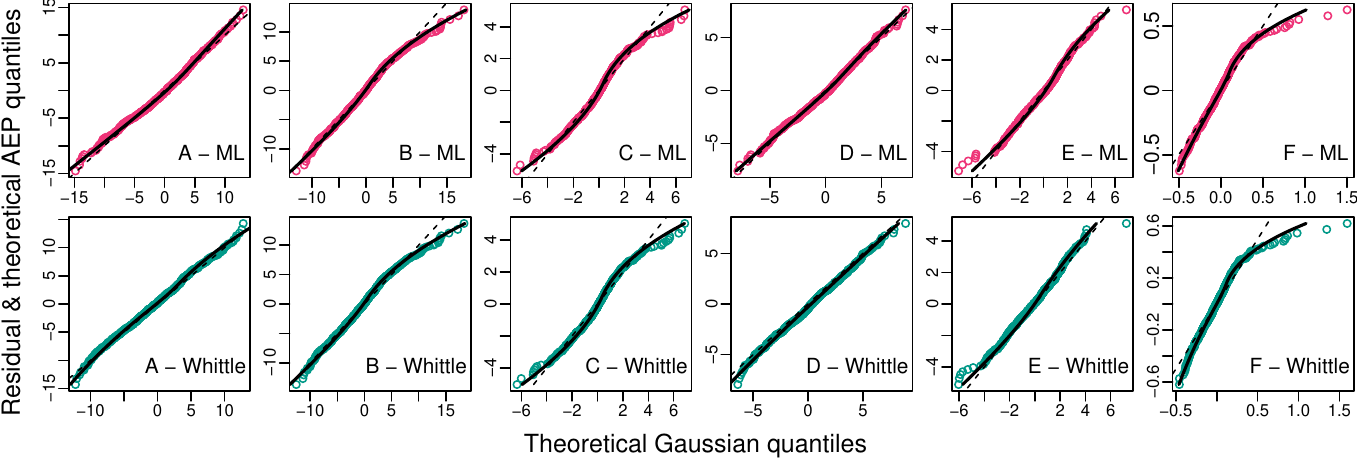}
 \caption{Q-Q plots for boreholes A--F of quantiles of residuals from mixed models estimated using maximum likelihood and Whittle (magenta and teal circles, respectively) as well as quantiles of AEP distributions fitted to the residuals (solid lines) against quantiles of the theoretical Gaussian distribution using parameter estimates from the mixed models (dashed lines).}
 \label{fig:qq_model}
\end{figure}
We now assess key model diagnostics of the error term. Figure~\ref{fig:acv_model} presents the empirical autocovariance of model residuals and autocovariance functions with estimated parameters. The covariance functions estimated using both maximum likelihood and Whittle were similar to the empirical ones for borehole C, and to a lesser extent for boreholes B and E, suggesting that for these boreholes the exponential covariance provided a relatively good fit to the data. For boreholes A, D, and F, there were substantial gaps between the empirical covariance and exponential covariance, suggesting that this covariance model was not flexible enough to describe the behaviour of the random error. Such gaps were lessened for boreholes D and F when a Mat\'ern covariance was estimated using Whittle. For borehole A, a good fit was achieved after estimating a model with Mat\'ern-periodic covariance. \par
Figure~\ref{fig:qq_model} presents Q-Q plots of model residuals against theoretical Gaussian quantiles showing clear departures from Gaussianity: in boreholes B, C and F, the right tails are heavy; in borehole C, the left tail is also heavy; and in borehole F, the left tail is light. The departures from Gaussianity are seen regardless of inference method used. This suggests that for boreholes B, C, and F, the key assumption of Gaussian maximum likelihood was violated. The figure also juxtaposes the empirical distributions of model residuals with AEP distributions fitted to the residuals. Using AEP improved the fit in all three cases (B, C and F). This suggests that for these three boreholes, the AEP distribution was a more appropriate model for the random error than a Gaussian distribution. We note that optimising the log-likelihood function for the full 17-parameter mixed model (including spline basis functions) with AEP-distributed errors required much longer computation than Whittle or Gaussian maximum likelihood, and optimisation algorithms converged to local minima returning no plausible estimates. We therefore conclude that the estimation of the full model using maximum likelihood is unfeasible.\par
Finally, we assessed the predictive power of the fitted models. We randomly removed and predicted (using Kriging) 12 monthly values using gaps of either: 12($\times 1$), 6($\times 2$), or 3($\times 4$) months of consecutively missing data. In each case the introduced gaps neighboured observed values. We repeated the procedure 250 times. Table~\ref{tab:RMSE} shows how using Whittle reduced the RMSE of predictions relative to Gaussian maximum likelihood. Whittle offered the greatest improvement for boreholes A and F, for which the model fit differed the most between Whittle and maximum likelihood (see Figure~\ref{fig:acv_model}). For borehole A, the yearly periodicity in the covariance of the model residuals corresponded to an especially large improvement for the 12-month gaps. Smaller improvements were also recorded for gaps of 3 and 6 months for boreholes B and D. For 12-month gaps for borehole D, however, the RMSE for Whittle was slightly higher than for maximum likelihood. The smallest differences of less than 1\% were recorded for boreholes C, E, and for 12-month gaps for borehole B, where the two methods performed similarly overall. This is likely because for these three boreholes, the advantage of using Whittle over maximum likelihood was the smallest in terms of model fit to the data (see Figure~\ref{fig:acv_model}). \par

\begin{table}[tb]
\centering
\caption{Percentage reduction of RMSE of missing value prediction for Whittle relative to Gaussian maximum likelihood for introduced gaps of different lengths. Higher values indicate stronger advantage of Whittle.}
\begin{tabular*}{\columnwidth}{@{\extracolsep\fill}lcccccc@{\extracolsep\fill}}
\toprule
& \multicolumn{6}{@{}c@{}}{Borehole} \\
\cline{2-7}
 Months missing     & A      & B     & C     & D      & E      & F      \\
 \midrule
12 months $\times$ 1 & 20.37\% & 0.92\% & 0.49\% & -1.15\% & 0.53\% & 7.39\% \\
6 months $\times$ 2  & 15.13\%  & 1.26\% & 0.67\% & 1.42\%  & -0.23\% & 8.19\%  \\
3 months $\times$ 4  & 6.60\%  & 2.85\% & 0.19\% & 3.06\%  & -0.92\% & 9.01\% \\
\hline
\end{tabular*}
\label{tab:RMSE}
\end{table}

In an analysis which we present in detail in Appendix~\ref{apx:identifiability}, we verified that the model parameters were identifiable. Considering evidence from both our application-motivated simulation study as well as from boreholes A--F, we confirmed that there were no identifiability issues between the fixed and random terms in practice. Notwithstanding this, with other applications or datasets the fixed term may operate on a similar timescale as the random term. In our mixed model setup, such examples could include precipitation and model random error (both stochastic) having similar covariance functions, or a spline basis function covering a time period of a length similar to the covariance function decay. While there was no evidence of these being a problem in our analysis, we encourage users to verify this in their application and with their dataset---either through similar analyses to those we present in the Appendix, or through expert prior knowledge of the timescales driving the constituent processes that form the mixed model. We also note that the parameters $\bgamma=(s,a)^T$ of the IRF are not generally identifiable, but our interest is in the resulting shapes of the IRFs rather than the parameter values themselves. While this lack of identifiability is therefore not a problem from the perspective of model interpretation, it may make numerical optimisation more difficult (independent of the estimation method used), so we reiterate our advice to repeat the optimisation with different starting values to help ensure the convergence is to the global minimum.

\section{Concluding remarks}\label{sec:conclusions}
This paper proposed a new Whittle-based procedure for the simultaneous estimation of mixed model parameters in time series with missing data. In a simulation study, we demonstrated that our proposed method offered an advantage over 2-stage estimation (fixed and random components estimated separately), leading to estimates and predictions closer to those from Gaussian maximum likelihood, which is known to have various desirable properties if the error term is indeed Gaussian. In the scenario with an AEP-distributed error term, however, we found that Whittle outperformed Gaussian maximum likelihood with improved estimates. Such robustness to non-Gaussianity was also found in \cite{jesus2017inference} and \cite{sykulski2019debiased}, but less so in the study of \cite{contreras2006note}. None of these studies explicitly compared with the Gaussian maximum likelihood estimates---our paper therefore provides a concrete example of Whittle-based inference outperforming Gaussian maximum likelihood under misspecification in a simulation study, as well as in application. \par
In application to GWLs, the computational time advantage of Whittle, through the use of FFTs, allowed us to estimate more complex covariance models than the exponential covariance (used in practice with maximum likelihood). This significantly improved model fit and predictions for certain time series, for which exponential covariance was insufficiently flexible. For boreholes for which the exponential covariance already provided a good fit, Whittle still led to similar predictions as maximum likelihood in less computational time. This is in agreement with the results of our simulation study, in which we estimated equally flexible models using both Whittle and maximum likelihood, and did not observe very large differences in their predictive ability when the maximum likelihood was correctly specified. \par
Looking at model residuals, we found evidence of non-Gaussianity in three GWL time series, which violates the assumptions of Gaussian maximum likelihood. We showed that the AEP distribution fitted the residuals better. Hence, Whittle estimation should be favoured for the presented data, as it does not require a marginal distribution of error to be specified and, in our simulation study, we specifically showed that it led to better estimates than Gaussian maximum likelihood when the errors were AEP-distributed. Investigating whether such robustness to non-Gaussianity offered by Whittle methods extends to other applications is an interesting avenue of future work.\par
The Whittle likelihood optimisation may be further improved by using the function's first and second derivatives with respect to the model parameters. For the mixed model used in this paper (with Mat\'ern covariance), our Whittle likelihood function is differentiable in all parameters and most derivatives can be written analytically (with the exception of the smoothness parameter $\nu$, for which the derivatives would need to be obtained numerically). Such implementations, however, are specific to each application and model used, so we focused our analyses using a generic optimisation approach which can be immediately applied to various data and models---specifically, we used the \texttt{optim} function in \texttt{R} with the BFGS algorithm, which utilises numerical approximation to derivatives.

\section*{Acknowledgements}
The simulation study was carried out using the high-performance computing resources at the Research Computing Service at Imperial College London, doi: \url{https://doi.org/10.14469/hpc/2232}.

\section*{CRediT author contribution statement}
Jakub J. Pypkowski: Conceptualisation \textit{(lead)}, Formal Analysis, Investigation, Methodology, Software, Visualisation, Writing --- original draft. Adam M. Sykulski: Conceptualisation \textit{(supporting)}, Supervision, Writing --- review \& editing. James S. Martin: Conceptualisation \textit{(supporting)}, Supervision, Writing --- review \& editing. Ben P. Marchant: Conceptualisation \textit{(supporting)}, Data Curation.

\section*{Funding}
This work was supported by the Department of Mathematics, Imperial College London, via a Roth Scholarship. Ben P. Marchant's contribution was funded by NERC programme (NE/W004984/1).

\section*{Software and code availability}
\textit{Upon the publication of the article, we will share our \texttt{R} code used for the simulation study and the real data analysis on GitHub.}

\appendix
\renewcommand{\thesection}{A.\arabic{section}}
\renewcommand{\thefigure}{A\arabic{figure}}
\renewcommand{\thetable}{A\arabic{table}}
\renewcommand{\theequation}{A\arabic{equation}}
\setcounter{section}{0}
\setcounter{figure}{0}
\setcounter{table}{0}
\setcounter{equation}{0}
\section{AEP distribution}\label{apx:AEP}
 In this section, we present key properties of the AEP distribution derived by \cite{Zhu2009AEP}, its simulation using the inversion method, and the likelihood function for a mixed model with AEP-distributed errors. Let $Z$ be an AEP-distributed random variable, and let $z$ be its realisation. The probability density function is given as
 \begin{equation}
 f(z)=\begin{cases}
 \frac{\varsigma}{\varsigma^*\sigma}K_{EP}(p_1)\exp\left( -\frac{1}{p_1}\left| \frac{z-\mu}{2\varsigma^*\sigma} \right|^{p_1} \right) &\text{if } z\leq\mu,\\
 \frac{1-\varsigma}{(1-\varsigma^*)\sigma}K_{EP}(p_2)\exp\left( -\frac{1}{p_2}\left| \frac{z-\mu}{2(1-\varsigma^*)\sigma} \right|^{p_2} \right) &\text{if } z>\mu,
 \end{cases}
 \end{equation}
 where $K_{EP}(p)= [2p^{1/p}\Gamma(1+1/p)]^{-1}$ is a normalising constant, and $\varsigma^* = \varsigma K_{EP}(p_1) / [\varsigma K_{EP}(p_1)+(1-\varsigma) K_{EP}(p_2)]$. The cumulative distribution function (cdf) is given as
 \begin{equation}
 F(z) = \begin{cases}
 \varsigma\left[1-G\left( \frac{1}{p_1}\left| \frac{z-\mu}{2\varsigma^*\sigma} \right|^{p_1} ; \frac{1}{p_1} \right) \right] &\text{if } z\leq\mu,\\
 \varsigma+ (1-\varsigma)G\left( \frac{1}{p_2}\left| \frac{z-\mu}{2(1-\varsigma^*)\sigma} \right|^{p_2} ; \frac{1}{p_2} \right)&\text{if } z>\mu,
 \end{cases}
 \end{equation}
 where $G(x;\lambda)$ is the one-parameter Gamma cdf. Lastly, the quantile function is given as
 \begin{equation}\label{eq:AEPquant}
 F^{-1}(\pi) = \begin{cases}
 \mu - 2\sigma\varsigma^*\left[ p_1G^{-1}\left( 1-\frac{\pi}{\varsigma}; \frac{1}{p_1} \right) \right]^{1/p_1} & \text{if } \pi \leq \varsigma, \\
 \mu+2\sigma(1-\varsigma^*) \left[ p_2G^{-1}\left( 1-\frac{1-\pi}{1-\varsigma}; \frac{1}{p_2} \right) \right]^{1/p_2} & \text{if } \pi > \varsigma,
 \end{cases} \quad \pi \in [0,1].
 \end{equation}
 Following \cite{Zhu2009AEP}, we note that the AEP distribution has finite fourth-order moments.
 \par
 To simulate AEP-distributed errors, we generate Gaussian errors with the Mat\'ern covariance function $c_{\balpha}(\tau)$ and convert them to AEP-distributed errors using the inversion method \cite[][p.51]{Copulas2006}, as described below. We omit details of the theory on copulas, and refer the interested reader to \cite{Copulas2006}. Let $\Phi(x)$ denote the standard normal cdf. The AEP-distributed errors are then obtained as follows.
 \begin{enumerate}
 \item Generate $\bepsilon = (\epsilon_1, \dots, \epsilon_n)$ from the zero-mean Gaussian process with Mat\'ern covariance $c_{\balpha}(\tau)$.
 \item Convert $\bepsilon$ to the uniform distribution on $(0,1)$ as $\bepsilon^{(U)}= (\epsilon_1^{(U)}, \dots, \epsilon_n^{(U)}) = (\Phi(\epsilon_1), \dots, \Phi(\epsilon_n))$.
 \item Convert $\bepsilon^{(U)}$ to the AEP distribution as $\bepsilon^{(AEP)}= (\epsilon_1^{(AEP)}, \dots, \epsilon_n^{(AEP)}) = (F^{-1}(\epsilon_1^{(U)}), \dots, F^{-1}(\epsilon_n^{(U)}))$ using the quantile function of the desired AEP distribution given in \eqref{eq:AEPquant}.
 \end{enumerate}\par
The log-likelihood function for a mixed model with AEP-distributed errors is specified as follows [\citealp{kazianka2010copula}; \citealp{MARCHANT2016}].
 \begin{equation}
 l_{AEP}(\balpha,\bvartheta, \btheta) = -\frac{1}{2}\log|C(\balpha)|+\frac{1}{2}\mathbf{a}^T_{\btheta}(I_n-C(\balpha)^{-1})\mathbf{a}_{\btheta} + \sum_{t=1}^n \log f[x_t- \phi_t(\bvartheta)],
 \end{equation}
 where $\mathbf{a}_{\btheta}=(a_{\btheta,1}, \dots, a_{\btheta,n})$ is obtained from errors $\bepsilon^{(AEP)}=\mathbf{x}-\bphi(\bvartheta)$ by reversing the inversion method, that is $a_{\btheta,t} = \Phi^{-1}\{F[x_t-\phi_t(\bvartheta)]\}$.

\section{Identifiability analysis}\label{apx:identifiability}
  In this section, we present our analysis of parameter identifiability in the simulated and real GWL data.\par
 \begin{figure}[!htb]
 \centering     
     \includegraphics[width=0.70\linewidth]{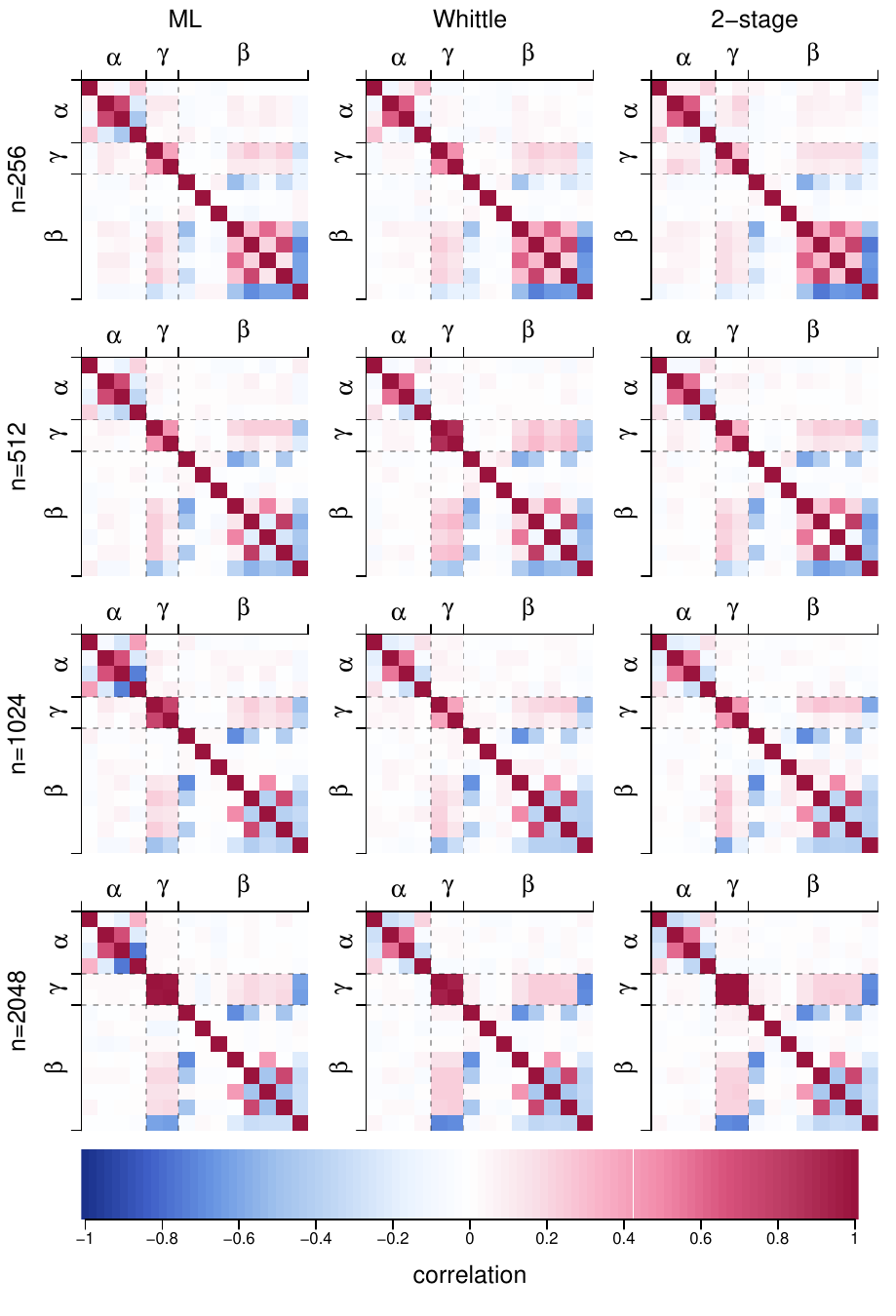}
\caption{Covariance matrices of parameter estimates in the standard mixed model scenario of the simulation study.}
 \label{fig:corr_est_Gauss}
\end{figure}
\begin{figure}[!htb]
 \centering     
     \includegraphics[width=0.70\linewidth]{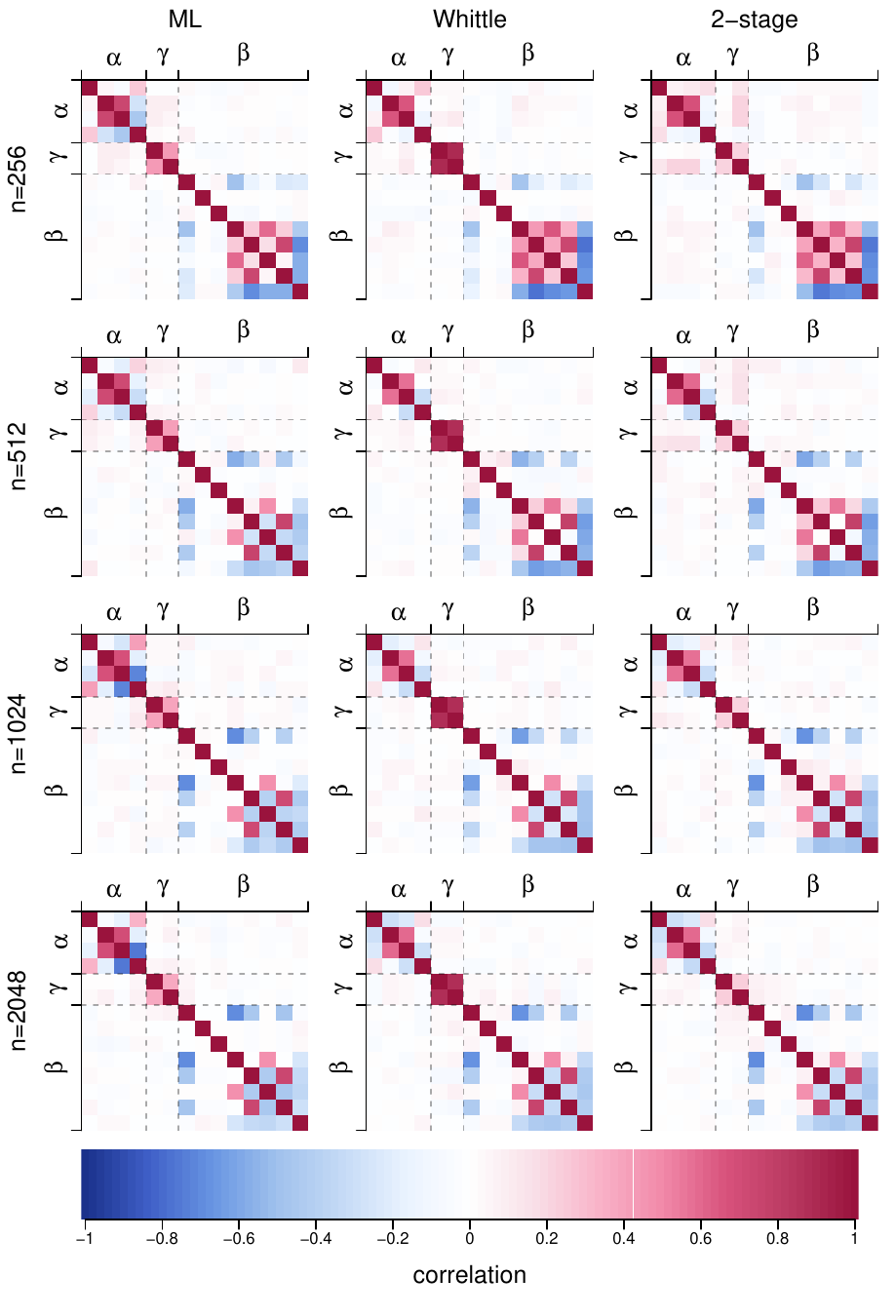}
 \caption{Covariance matrices of parameter estimates in the fully random model scenario of the simulation study.}
 \label{fig:corr_est_Zero}
\end{figure}
\begin{figure}[!htb]
 \centering     
     \includegraphics[width=0.87\linewidth]{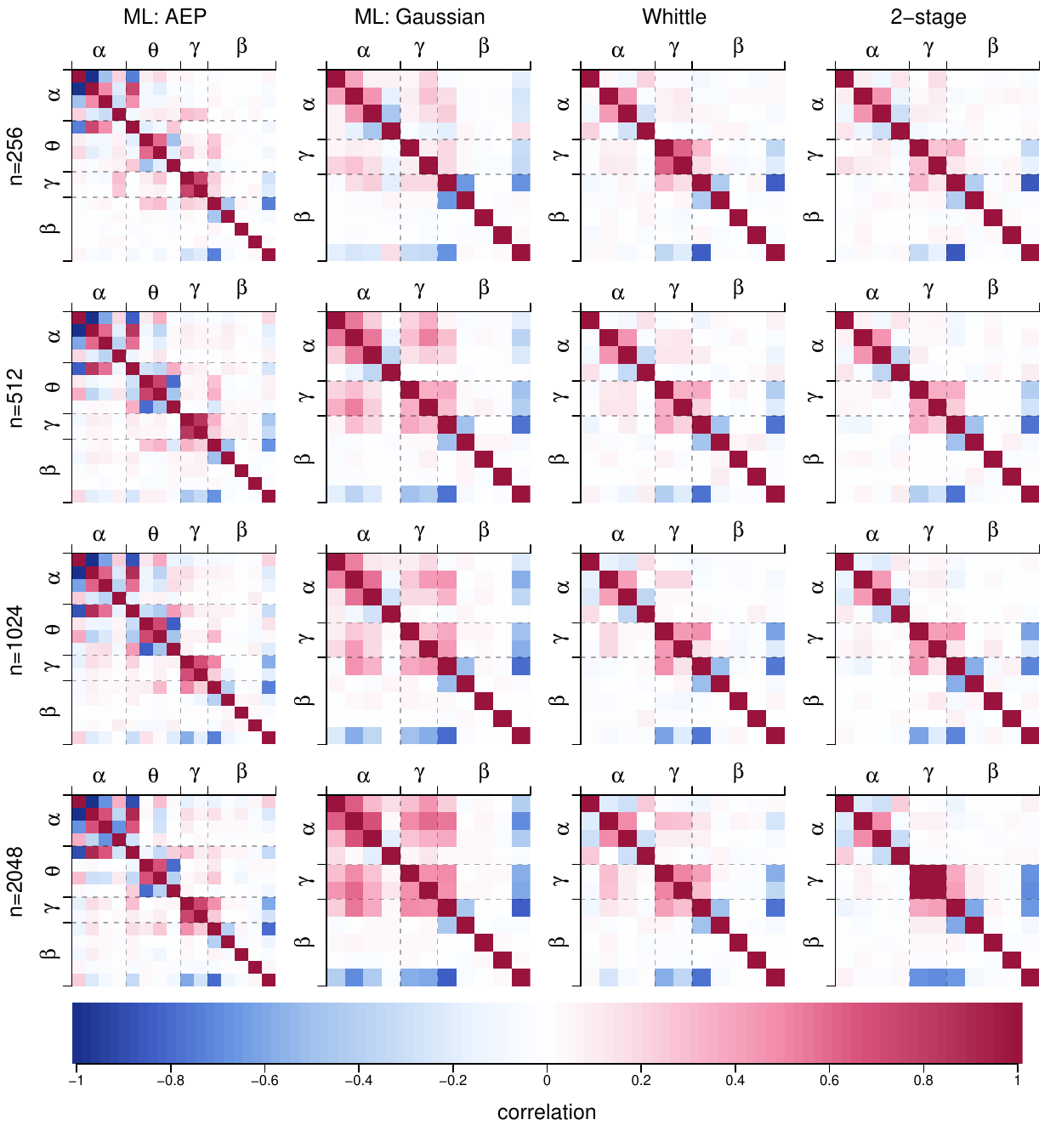}
 \caption{Covariance matrices of parameter estimates in the simulation study scenario with AEP-distributed errors.}
 \label{fig:corr_est_AEP}
\end{figure}
We first consider the correlations between parameter estimates for all scenarios of our simulation study, which are presented in Figures~\ref{fig:corr_est_Gauss}--\ref{fig:corr_est_AEP}. Strong correlation (either positive or negative) between two parameter estimates indicates that the parameters may be difficult to identify. There are no signs of strong correlation between $\bbeta$ parameters (linear fixed effects) and $\balpha$ parameters (random effects), apart from the misspecified Gaussian maximum likelihood for the model with AEP-distributed errors (Figure~\ref{fig:corr_est_AEP}, middle column), which is likely an impact of misspecification. Some weak correlation is also found in Figure~\ref{fig:corr_est_AEP} (left column) between parameters $\btheta$ (the AEP marginal distribution parameters) and the intercept for the correctly specified maximum likelihood with AEP-distributed errors. Such correlation is not of concern, as the role of the intercept is solely to offset for the other components of the model.\par
\begin{figure}[t]
 \centering     
     \includegraphics[width=0.78\linewidth]{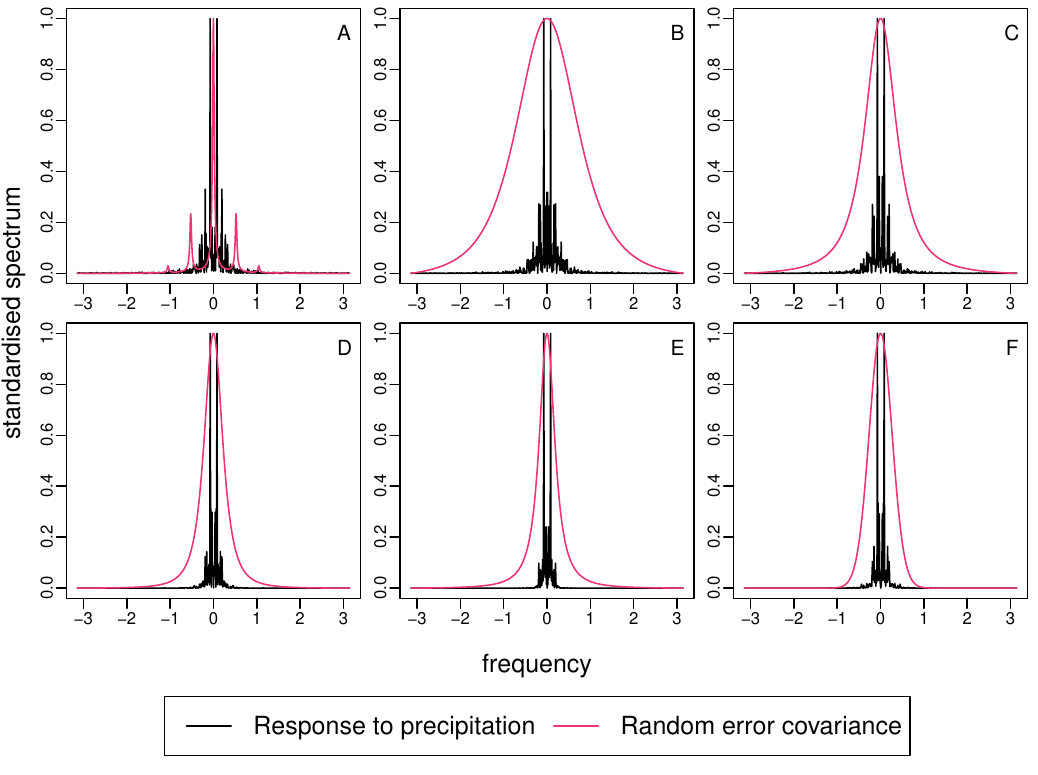}
 \caption{Comparison of relative spectra, given as $[f(\omega)-\min_{\omega\in(-\pi,\pi)}f(\omega)]/[\max_{\omega\in(-\pi,\pi)}f(\omega)-\min_{\omega\in(-\pi,\pi)}f(\omega)]$, of the estimated response to precipitation and random error covariance for the six boreholes A--F.}
 \label{fig:spectrum_IRF_cov}
 \end{figure}

There is some weak correlation between parameters $\bgamma$ and $\balpha$ in Figures~\ref{fig:corr_est_Gauss}~and~\ref{fig:corr_est_AEP}, across all estimation methods. This can be explained by the fact that the IRFs are related to the precipitation (an external variable), which is generated by a random process with temporal correlation---similarly to the random error. The parameters are identifiable if these processes operate on different timescales. To check if they do in practice, we consider power spectral densities of the response to precipitation and the random error estimated for real GWL data from boreholes~A--F, which are displayed in Figure~\ref{fig:spectrum_IRF_cov}. The densities have different shapes for all boreholes, confirming that they operate on different timescales. For borehole A, the spectrum of the response to precipitation decays slower than the central peak of the spectrum of the random error, and the latter distinguishes itself by another peak at $\pi/12$, which corresponds to the 12-month periodicity. For boreholes B--F, the spectra of the random error decay slower than those of the response to precipitation. These differences guarantee that the two components are identifiable in the real data used in the paper. In other applications, should one predictor and the random error operate on similar timescales, the identifiability may be compromised.\par

Similarly, the identifiability between the spline components (i.e. temporary trends) and temporal dependence of the error term may be a problem if these operate on similar timescales. In boreholes A--F, however, we found that the covariance decays substantially in approx. 5 years for borehole A, $<2$ years for borehole E, and $<1$ year for all other boreholes. These periods are substantially shorter than the span of one spline basis function used in the model, which is 15 years. That is, in practice these operate on different timescales. We note that such consideration also justifies the selection of the Mat\'ern covariance for modelling, as it is a common model for short-memory processes. Such processes can still exhibit long-term dependence (while retaining short-memory), but such behaviour is not expected in our application (and the Mat\'ern covariance is not usually used in such instances). We advise that users of similar mixed models refer to expert knowledge to establish the possible ranges of temporal dependence to ensure that identifiability is not compromised in their application.\par
For borehole A, identifiability between the fixed seasonal effect and the random term with the Mat\'ern-periodic covariance is guaranteed by the following differences seen in Fig~\ref{fig:spectrum_IRF_cov}. The spectrum of the Mat\'ern-periodic covariance contains a central peak at 0 and smaller peaks at $\pm \pi/12$ (and even smaller at its multiples), but is nonzero on $[-\pi, \pi]$. This can be contrasted with the spectrum of the seasonal component of the fixed term whose spectrum consists of only Dirac deltas at $\pm \pi/12$. The spectrum is 0 at all other frequencies and there is no central peak.\par

\begin{figure}[!htb]
 \centering     
     \includegraphics[width=0.78\linewidth]{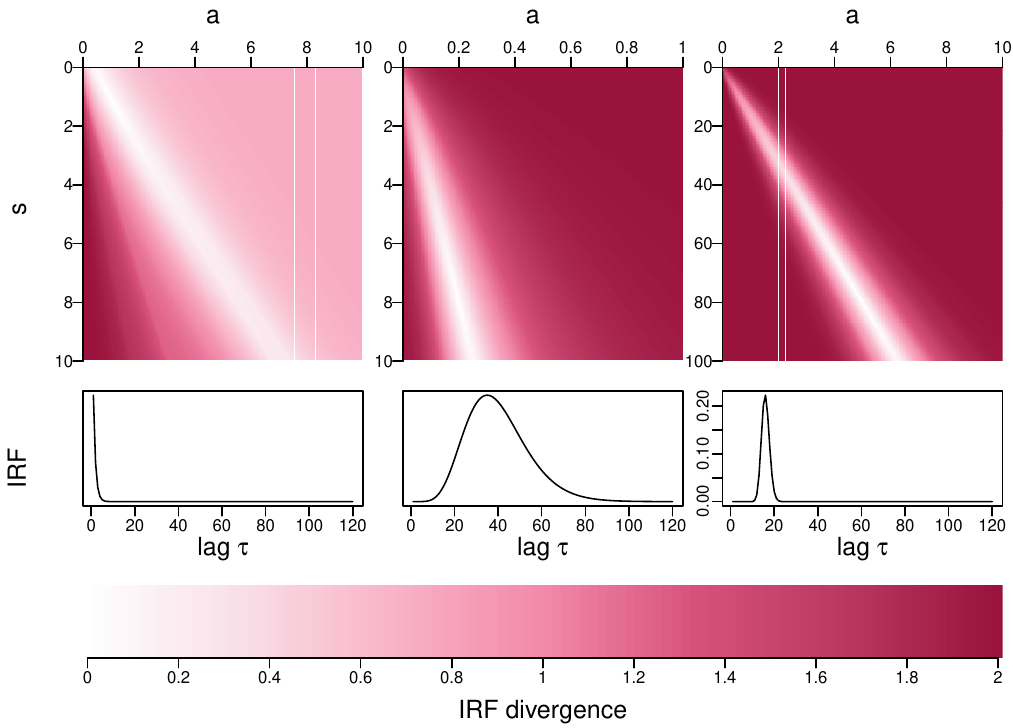}
 \caption{Divergence of IRFs produced by different pairs of parameters $(s,a)$ from IRFs produced by the following pairs, respectively: $(1,1)$;  $(8,0.2)$; $(80,5)$. The shapes of the corresponding IRFs are presented below the divergence plots.} 
 \label{fig:IRF_div}
\end{figure}

Lastly, we recall that there is no practical interest in finding specific values of parameters $\bgamma=(s,a)^T$ as they are not interpretable, and we are only interested in IRFs the estimates produce. Using the divergence metric $\text{Div}[IRF(s,a; \hat s, \hat a)] =\sum_{\tau=0}^{n_{IRF}-1} |IRF(\tau; \hat s, \hat a) - IRF(\tau; s, a)|$, we show in Figure~\ref{fig:IRF_div} that IRFs hold resemblance along certain valleys in the parameter space. Because of this, errors in the parameter estimates are likely to be positively correlated, i.e. if we overestimate $s$, we are likely also overestimating $a$ etc., as we had shown in Figures~\ref{fig:corr_est_Gauss}~\&~\ref{fig:corr_est_AEP}, where strong correlation is found between estimates of $\bgamma=(s,a)^T$. We recall from our simulation study that the average IRF divergence decreases with sample size, as expected. Therefore, the correlation between the estimates, which remains strong for large samples, does not impede the observed decrease in the IRF divergence.

\begin{thebibliography}{59}
\providecommand{\natexlab}[1]{#1}
\providecommand{\url}[1]{\texttt{#1}}
\expandafter\ifx\csname urlstyle\endcsname\relax
  \providecommand{\doi}[1]{doi: #1}\else
  \providecommand{\doi}{doi: \begingroup \urlstyle{rm}\Url}\fi

\bibitem[Aigrain and Foreman-Mackey(2023)]{aigrain2023astrostat}
S.~Aigrain and D.~Foreman-Mackey.
\newblock Gaussian process regression for astronomical time series.
\newblock \emph{Annual Review of Astronomy and Astrophysics}, 61\penalty0 (1):\penalty0 329--371, 2023.
\newblock \doi{https://doi.org/10.1146/annurev-astro-052920-103508}.

\bibitem[Alman and Williams(2024)]{alman2024matrixInv}
J.~Alman and V.~V. Williams.
\newblock A refined laser method and faster matrix multiplication.
\newblock \emph{TheoretiCS}, 3, 2024.
\newblock \doi{https://doi.org/10.46298/theoretics.24.21}.

\bibitem[Baltagi et~al.(2025)Baltagi, Bresson, and Etienne]{Baltagi2025}
B.~H. Baltagi, G.~Bresson, and J.-M. Etienne.
\newblock \emph{Estimation of Serially Correlated Error Components Models Using Whittle's Approximate Maximum Likelihood Method}, pages 337--365.
\newblock Springer Nature Switzerland, Cham, 2025.
\newblock \doi{https://doi.org/10.1007/978-3-031-92699-0_12}.

\bibitem[Bellio et~al.(2025)Bellio, Ghosh, Owen, and Varin]{Bellio2025}
R.~Bellio, S.~Ghosh, A.~B. Owen, and C.~Varin.
\newblock Consistent and scalable composite likelihood estimation of probit models with crossed random effects.
\newblock \emph{Biometrika}, 112\penalty0 (3):\penalty0 asaf037, 2025.
\newblock \doi{https://doi.org/10.1093/biomet/asaf037}.

\bibitem[Bertolacci et~al.(2022)Bertolacci, Rosen, Cripps, and Cripps]{Bertolacci2019}
M.~Bertolacci, O.~Rosen, E.~Cripps, and S.~Cripps.
\newblock Adaptspec-x: Covariate-dependent spectral modeling of multiple nonstationary time series.
\newblock \emph{Journal of Computational and Graphical Statistics}, 31\penalty0 (2):\penalty0 436--454, 2022.
\newblock \doi{https://doi.org/10.1080/10618600.2021.2000870}.

\bibitem[Bloomfield and Marchant(2013)]{BloomfieldSGI2013}
J.~P. Bloomfield and B.~P. Marchant.
\newblock Analysis of groundwater drought building on the standardised precipitation index approach.
\newblock \emph{Hydrology and Earth System Sciences}, 17\penalty0 (12):\penalty0 4769--4787, 2013.
\newblock \doi{https://doi.org/10.5194/hess-17-4769-2013}.

\bibitem[Contreras-Crist{\'a}n et~al.(2006)Contreras-Crist{\'a}n, Guti{\'e}rrez-Pe{\~n}a, and Walker]{contreras2006note}
A.~Contreras-Crist{\'a}n, E.~Guti{\'e}rrez-Pe{\~n}a, and S.~G. Walker.
\newblock A note on {W}hittle's likelihood.
\newblock \emph{Communications in Statistics-Simulation and Computation}, 35\penalty0 (4):\penalty0 857--875, 2006.
\newblock \doi{https://doi.org/10.1080/03610910600880203}.

\bibitem[Coutinho et~al.(2023)Coutinho, Madeira, da~Silva, de~Oliveira, and Delgado]{CoutinhoEluãRamos2023AoaC}
E.~R. Coutinho, J.~G.~F. Madeira, R.~M. da~Silva, E.~M. de~Oliveira, and A.~R.~S. Delgado.
\newblock Application of a computational hybrid model to estimate and filling gaps for meteorological time series.
\newblock \emph{Revista Brasileira de Meteorologia}, 38:\penalty0 1, 2023.
\newblock \doi{https://doi.org/10.1590/0102-778638220030}.

\bibitem[Dahlhaus(1988)]{dahlhaus1988small}
R.~Dahlhaus.
\newblock Small sample effects in time series analysis: a new asymptotic theory and a new estimate.
\newblock \emph{The Annals of Statistics}, 16\penalty0 (2):\penalty0 808--841, 1988.
\newblock \doi{https://doi.org/10.1214/aos/1176350838}.

\bibitem[Dahlhaus(2000)]{dahlhaus2000likelihood}
R.~Dahlhaus.
\newblock A likelihood approximation for locally stationary processes.
\newblock \emph{The Annals of Statistics}, 28\penalty0 (6):\penalty0 1762--1794, 2000.
\newblock \doi{https://doi.org/10.1214/aos/1015957480}.

\bibitem[D'Errico(2012)]{fminsearchbnd}
J.~D'Errico.
\newblock fminsearchbnd, fminsearchcon.
\newblock MATLAB Central File Exchange, 2012.
\newblock URL \url{https://www.mathworks.com/matlabcentral/fileexchange/8277-fminsearchbnd-fminsearchcon}.
\newblock Accessed: 2 May 2013.

\bibitem[ElHaj and Alshamsi(2025)]{elhaj2025geotemporal}
K.~ElHaj and D.~Alshamsi.
\newblock Geotemporal clustering for aquifer delineation: a big data approach to synchronizing and analyzing variable-length groundwater time series.
\newblock \emph{Journal of Big Data}, 12\penalty0 (1):\penalty0 25, 2025.
\newblock \doi{https://doi.org/10.1186/s40537-025-01060-6}.

\bibitem[Ghandwani et~al.(2025)Ghandwani, Ghosh, Hastie, and Owen]{Ghandwani2025}
D.~Ghandwani, S.~Ghosh, T.~Hastie, and A.~Owen.
\newblock Scalable solutions for crossed random-effect models with random slopes.
\newblock \emph{Electronic Journal of Statistics}, 19\penalty0 (2), 2025.
\newblock ISSN 1935-7524.
\newblock \doi{https://doi.org/10.1214/25-EJS2461}.

\bibitem[Gholizadeh et~al.(2023)Gholizadeh, Zhang, Frame, Gu, and Green]{Gholizadeh2023}
H.~Gholizadeh, Y.~Zhang, J.~Frame, X.~Gu, and C.~T. Green.
\newblock Long short-term memory models to quantify long-term evolution of streamflow discharge and groundwater depth in {A}labama.
\newblock \emph{Science of The Total Environment}, 901:\penalty0 165884, 2023.
\newblock \doi{https://doi.org/10.1016/j.scitotenv.2023.165884}.

\bibitem[Ghosh et~al.(2022)Ghosh, Hastie, and Owen]{Ghosh2022}
S.~Ghosh, T.~Hastie, and A.~B. Owen.
\newblock Backfitting for large scale crossed random effects regressions.
\newblock \emph{The Annals of Statistics}, 50\penalty0 (1):\penalty0 560--583, 2022.
\newblock ISSN 0090-5364.
\newblock \doi{https://doi.org/10.1214/21-AOS2121}.

\bibitem[Goodwin et~al.(2025)Goodwin, Quiroz, and Kohn]{goodwin2025}
T.~Goodwin, M.~Quiroz, and R.~Kohn.
\newblock Dynamic linear regression models for forecasting time series with semi long memory errors.
\newblock arXiv:\penalty0 2408.09096v2 [stat.ME], 2025.
\newblock \doi{https://doi.org/10.48550/arXiv.2408.09096}.

\bibitem[Goplerud et~al.(2025)Goplerud, Papaspiliopoulos, and Zanella]{Goplerud2025}
M.~Goplerud, O.~Papaspiliopoulos, and G.~Zanella.
\newblock Partially factorized variational inference for high-dimensional mixed models.
\newblock \emph{Biometrika}, 112\penalty0 (2):\penalty0 asae067, 2025.
\newblock \doi{https://doi.org/10.1093/biomet/asae067}.

\bibitem[Guillaumin et~al.(2017)Guillaumin, Sykulski, Olhede, Early, and Lilly]{guillaumin2017modulated}
A.~P. Guillaumin, A.~M. Sykulski, S.~C. Olhede, J.~J. Early, and J.~M. Lilly.
\newblock Analysis of non-stationary modulated time series with applications to oceanographic surface flow measurements.
\newblock \emph{Journal of Time Series Analysis}, 38\penalty0 (5):\penalty0 668--710, 2017.
\newblock \doi{https://doi.org/10.1111/jtsa.12244}.

\bibitem[Guillaumin et~al.(2022)Guillaumin, Sykulski, Olhede, and Simons]{Guillaumin2022DebiasedSpatial}
A.~P. Guillaumin, A.~M. Sykulski, S.~C. Olhede, and F.~J. Simons.
\newblock The debiased spatial {W}hittle likelihood.
\newblock \emph{Journal of the Royal Statistical Society Series B: Statistical Methodology}, 84\penalty0 (4):\penalty0 1526--1557, 07 2022.
\newblock \doi{https://doi.org/10.1111/rssb.12539}.

\bibitem[Hiben et~al.(2024)Hiben, Awoke, and Ashenafi]{hiben2024estimation}
M.~G. Hiben, A.~G. Awoke, and A.~A. Ashenafi.
\newblock Estimation of rainfall and streamflow missing data under uncertainty for {Nile} basin headwaters: the case of {Ghba} catchments.
\newblock \emph{Journal of Applied Water Engineering and Research}, 12\penalty0 (2):\penalty0 119--133, 2024.
\newblock \doi{https://doi.org/10.1080/23249676.2023.2230892}.

\bibitem[Huang et~al.(2016)Huang, Xia, and Qin]{Huang2016}
L.~Huang, Y.~Xia, and X.~Qin.
\newblock Estimation of semivarying coefficient time series models with {ARMA} errors.
\newblock \emph{The Annals of Statistics}, 44\penalty0 (4):\penalty0 1618--1660, 2016.
\newblock URL \url{http://www.jstor.org/stable/43974727}.

\bibitem[Ivanov et~al.(2020)Ivanov, Leonenko, and Orlovskyi]{Ivanov2020Levy}
A.~V. Ivanov, N.~N. Leonenko, and I.~V. Orlovskyi.
\newblock On the {W}hittle estimator for linear random noise spectral density parameter in continuous-time nonlinear regression models.
\newblock \emph{Statistical Inference for Stochastic Processes}, 23\penalty0 (1):\penalty0 129--169, 2020.
\newblock \doi{https://doi.org/10.1007/s11203-019-09206-z}.

\bibitem[Ivanov and Prykhod’ko(2016)]{Ivanov2016Gauss}
O.~V. Ivanov and V.~V. Prykhod’ko.
\newblock On the {Whittle} estimator of the parameter of spectral density of random noise in the nonlinear regression model.
\newblock \emph{Ukrainian Mathematical Journal}, 67\penalty0 (8):\penalty0 1183--1203, 2016.
\newblock \doi{https://doi.org/10.1007/s11253-016-1145-1}.

\bibitem[Jesus and Chandler(2017)]{jesus2017inference}
J.~Jesus and R.~E. Chandler.
\newblock Inference with the {W}hittle likelihood: A tractable approach using estimating functions.
\newblock \emph{Journal of Time Series Analysis}, 38\penalty0 (2):\penalty0 204--224, 2017.
\newblock \doi{https://doi.org/10.1111/jtsa.12225}.

\bibitem[Kazianka and Pilz(2010)]{kazianka2010copula}
H.~Kazianka and J.~Pilz.
\newblock Copula-based geostatistical modeling of continuous and discrete data including covariates.
\newblock \emph{Stochastic Environmental Research and Risk Assessment}, 24\penalty0 (5):\penalty0 661--673, 2010.
\newblock \doi{https://doi.org/10.1007/s00477-009-0353-8}.

\bibitem[Koul and Surgailis(2008)]{koul2008testing}
H.~L. Koul and D.~Surgailis.
\newblock Testing a sub-hypothesis in linear regression models with long memory covariates and errors.
\newblock \emph{Applications of Mathematics}, 53\penalty0 (3):\penalty0 235--248, 2008.
\newblock \doi{https://doi.org/10.1007/s10492-008-0007-z}.

\bibitem[Li et~al.(2021)Li, Rosen, Ferrarelli, and Krafty]{Li2021}
Z.~Li, O.~Rosen, F.~Ferrarelli, and R.~T. Krafty.
\newblock Adaptive {Bayesian} spectral analysis of high-dimensional nonstationary time series.
\newblock \emph{Journal of Computational and Graphical Statistics}, 30\penalty0 (3):\penalty0 794--807, 2021.
\newblock \doi{https://doi.org/10.1080/10618600.2020.1868305}.

\bibitem[Little and Rubin(2020)]{missingData_txtbook}
R.~J.~A. Little and D.~B. Rubin.
\newblock \emph{Statistical analysis with missing data}.
\newblock Wiley series in probability and statistics. Wiley, Hoboken, NJ, third edition, 2020.

\bibitem[Loya et~al.(2025)Loya, Kalantzis, Cooper, and Palamara]{LoyaHrushikesh2025Asvi}
H.~Loya, G.~Kalantzis, F.~Cooper, and P.~F. Palamara.
\newblock A scalable variational inference approach for increased mixed-model association power.
\newblock \emph{Nature Genetics}, 57\penalty0 (2):\penalty0 461--468, 2025.
\newblock \doi{https://doi.org/10.1038/s41588-024-02044-7}.

\bibitem[Mackay et~al.(2014)Mackay, Jackson, and Wang]{MACKAY2014}
J.~Mackay, C.~Jackson, and L.~Wang.
\newblock A lumped conceptual model to simulate groundwater level time-series.
\newblock \emph{Environmental Modelling \& Software}, 61:\penalty0 229--245, 2014.
\newblock \doi{https://doi.org/10.1016/j.envsoft.2014.06.003}.

\bibitem[Marchant and Bloomfield(2018)]{MARCHANT2018}
B.~Marchant and J.~Bloomfield.
\newblock Spatio-temporal modelling of the status of groundwater droughts.
\newblock \emph{Journal of Hydrology}, 564:\penalty0 397--413, 2018.
\newblock \doi{https://doi.org/10.1016/j.jhydrol.2018.07.009}.

\bibitem[Marchant et~al.(2016)Marchant, Mackay, and Bloomfield]{MARCHANT2016}
B.~Marchant, J.~Mackay, and J.~Bloomfield.
\newblock Quantifying uncertainty in predictions of groundwater levels using formal likelihood methods.
\newblock \emph{Journal of Hydrology}, 540:\penalty0 699--711, 2016.
\newblock \doi{https://doi.org/10.1016/j.jhydrol.2016.06.014}.

\bibitem[Marchant et~al.(2022)Marchant, Cuba, Brauns, and Bloomfield]{marchant2022temporal}
B.~Marchant, D.~Cuba, B.~Brauns, and J.~Bloomfield.
\newblock Temporal interpolation of groundwater level hydrographs for regional drought analysis using mixed models.
\newblock \emph{Hydrogeology Journal}, 30\penalty0 (6):\penalty0 1801--1817, 2022.
\newblock \doi{https://doi.org/10.1007/s10040-022-02528-y}.

\bibitem[Myeni et~al.(2022)Myeni, Moeletsi, and Clulow]{MyeniL2022Davo}
L.~Myeni, M.~Moeletsi, and A.~Clulow.
\newblock Development and validation of an operational multi-layered model for estimation of soil moisture at point-scale in {South} {Africa}.
\newblock \emph{South African Journal of Plant and Soil}, 39\penalty0 (1):\penalty0 28--40, 2022.
\newblock \doi{https://doi.org/10.1080/02571862.2021.1970832}.

\bibitem[Nelsen(2006)]{Copulas2006}
R.~B. Nelsen.
\newblock \emph{An Introduction to Copulas}.
\newblock Springer Series in Statistics. Springer New York, New York, NY, 2nd edition, 2006.
\newblock \doi{https://doi.org/10.1007/0-387-28678-0}.

\bibitem[Neumann et~al.(2003)Neumann, Brown, Smedley, and Besien]{Neumann2003}
I.~Neumann, S.~Brown, P.~L. Smedley, and T.~Besien.
\newblock {Baseline Report Series: 7. The Great and the Inferior Oolite of the Cotswold district}.
\newblock British Geological Survey Commissioned Report CR/03/202N, 2003.
\newblock URL \url{https://nora.nerc.ac.uk/id/eprint/3572}.

\bibitem[Papaspiliopoulos et~al.(2020)Papaspiliopoulos, Roberts, and Zanella]{Papaspiliopoulos2020}
O.~Papaspiliopoulos, G.~O. Roberts, and G.~Zanella.
\newblock Scalable inference for crossed random effects models.
\newblock \emph{Biometrika}, 107\penalty0 (1):\penalty0 25--40, 2020.
\newblock \doi{https://doi.org/10.1093/biomet/asz058}.

\bibitem[Percival and Walden(1993)]{walden1993}
D.~B. Percival and A.~T. Walden.
\newblock \emph{Spectral Analysis for Physical Applications: Multitaper and Conventional Univariate Techniques}.
\newblock Cambridge University Press, Cambridge, 1993.
\newblock \doi{https://doi.org/10.1017/CBO9780511622762}.

\bibitem[Peterson and Western(2014)]{peterson2014TFN}
T.~Peterson and A.~Western.
\newblock Nonlinear time-series modeling of unconfined groundwater head.
\newblock \emph{Water Resources Research}, 50\penalty0 (10):\penalty0 8330--8355, 2014.
\newblock \doi{https://doi.org/10.1002/2013WR014800}.

\bibitem[Peterson and Western(2018)]{peterson2018interpolation}
T.~J. Peterson and A.~W. Western.
\newblock Statistical interpolation of groundwater hydrographs.
\newblock \emph{Water Resources Research}, 54\penalty0 (7):\penalty0 4663--4680, 2018.
\newblock \doi{https://doi.org/10.1029/2017WR021838}.

\bibitem[Rahman et~al.(2024)Rahman, Babla, Sultana, Saba, and Hoque]{rahman2024estimation}
S.~M. Rahman, M.~A.~H. Babla, R.~Sultana, S.~Saba, and A.~Hoque.
\newblock Estimation of missing daily temperature and rainfall for longer durations at {Hatiya} and {Sandwip} islands in the {Bay} of {Bengal}.
\newblock \emph{Journal of Earth System Science}, 133\penalty0 (2):\penalty0 109, 2024.
\newblock \doi{https://doi.org/10.1007/s12040-024-02318-y}.

\bibitem[Rosen et~al.(2012)Rosen, Wood, and Stoffer]{Rosen2012}
O.~Rosen, S.~Wood, and D.~S. Stoffer.
\newblock {AdaptSPEC}: adaptive spectral estimation for nonstationary time series.
\newblock \emph{Journal of the American Statistical Association}, 107\penalty0 (500):\penalty0 1575--1589, 2012.
\newblock \doi{https://doi.org/10.1080/01621459.2012.716340}.

\bibitem[Shabalala et~al.(2019)Shabalala, Moeletsi, Tongwane, and Mazibuko]{shabalala2019evaluation}
Z.~P. Shabalala, M.~E. Moeletsi, M.~I. Tongwane, and S.~M. Mazibuko.
\newblock Evaluation of infilling methods for time series of daily temperature data: Case study of {Limpopo} {Province}, {South} {Africa}.
\newblock \emph{Climate}, 7\penalty0 (7):\penalty0 86, 2019.
\newblock \doi{https://doi.org/10.3390/cli7070086}.

\bibitem[Sykulski et~al.(2016)Sykulski, Olhede, Lilly, and Danioux]{Sykulski2016Drifters}
A.~M. Sykulski, S.~C. Olhede, J.~M. Lilly, and E.~Danioux.
\newblock Lagrangian time series models for ocean surface drifter trajectories.
\newblock \emph{Journal of the Royal Statistical Society Series C: Applied Statistics}, 65\penalty0 (1):\penalty0 29--50, 01 2016.
\newblock \doi{https://doi.org/10.1111/rssc.12112}.

\bibitem[Sykulski et~al.(2019)Sykulski, Olhede, Guillaumin, Lilly, and Early]{sykulski2019debiased}
A.~M. Sykulski, S.~C. Olhede, A.~P. Guillaumin, J.~M. Lilly, and J.~J. Early.
\newblock The debiased {Whittle} likelihood.
\newblock \emph{Biometrika}, 106\penalty0 (2):\penalty0 251--266, 02 2019.
\newblock \doi{https://doi.org/10.1093/biomet/asy071}.

\bibitem[Tanguy et~al.(2021)Tanguy, Dixon, Prosdocimi, Morris, and Keller]{rain_data}
M.~Tanguy, H.~Dixon, I.~Prosdocimi, D.~Morris, and V.~Keller.
\newblock \emph{Gridded estimates of daily and monthly areal rainfall for the United Kingdom (1890-2019) {[CEH-GEAR]}}.
\newblock NERC EDS Environmental Information Data Centre, 2021.
\newblock \doi{https://doi.org/10.5285/dbf13dd5-90cd-457a-a986-f2f9dd97e93c}.

\bibitem[Trench(1964)]{trench1964toeplitz}
W.~F. Trench.
\newblock An algorithm for the inversion of finite toeplitz matrices.
\newblock \emph{Journal of the Society for Industrial and Applied Mathematics}, 12\penalty0 (3):\penalty0 515--522, 1964.
\newblock \doi{https://doi.org/10.1137/0112045}.

\bibitem[van Lanen et~al.(2016)van Lanen, Laaha, Kingston, Gauster, Ionita, Vidal, Vlnas, Tallaksen, Stahl, Hannaford, Delus, Fendekova, Mediero, Prudhomme, Rets, Romanowicz, Gailliez, Wong, Adler, Blauhut, Caillouet, Chelcea, Frolova, Gudmundsson, Hanel, Haslinger, Kireeva, Osuch, Sauquet, Stagge, and Van~Loon]{vanLanen2016hydrology}
H.~A. van Lanen, G.~Laaha, D.~G. Kingston, T.~Gauster, M.~Ionita, J.-P. Vidal, R.~Vlnas, L.~M. Tallaksen, K.~Stahl, J.~Hannaford, C.~Delus, M.~Fendekova, L.~Mediero, C.~Prudhomme, E.~Rets, R.~J. Romanowicz, S.~Gailliez, W.~K. Wong, M.-J. Adler, V.~Blauhut, L.~Caillouet, S.~Chelcea, N.~Frolova, L.~Gudmundsson, M.~Hanel, K.~Haslinger, M.~Kireeva, M.~Osuch, E.~Sauquet, J.~H. Stagge, and A.~F. Van~Loon.
\newblock Hydrology needed to manage droughts: the 2015 {European} case.
\newblock \emph{Hydrological Processes}, 30\penalty0 (17):\penalty0 3097--3104, 2016.
\newblock \doi{https://doi.org/10.1002/hyp.10838}.

\bibitem[van Lieshout(2019)]{van2019theory}
M.~N.~M. van Lieshout.
\newblock \emph{Theory of spatial statistics: A concise introduction}.
\newblock Chapman and Hall/CRC, 2019.
\newblock \doi{https://doi.org/10.1201/9780429052866}.

\bibitem[von Asmuth and Bierkens(2005)]{vonAsmuth2005modeling}
J.~R. von Asmuth and M.~F. Bierkens.
\newblock Modeling irregularly spaced residual series as a continuous stochastic process.
\newblock \emph{Water Resources Research}, 41\penalty0 (12), 2005.
\newblock \doi{https://doi.org/10.1029/2004WR003726}.

\bibitem[von Asmuth et~al.(2002)von Asmuth, Bierkens, and Maas]{vonAsmuth2002IRF}
J.~R. von Asmuth, M.~F. Bierkens, and K.~Maas.
\newblock Transfer function-noise modeling in continuous time using predefined impulse response functions.
\newblock \emph{Water Resources Research}, 38\penalty0 (12):\penalty0 23-1--23-12, 2002.
\newblock \doi{https://doi.org/10.1029/2001WR001136}.

\bibitem[Vu et~al.(2021)Vu, Jardani, Massei, and Fournier]{Vu2021LSTM}
M.~Vu, A.~Jardani, N.~Massei, and M.~Fournier.
\newblock Reconstruction of missing groundwater level data by using long short-term memory {(LSTM)} deep neural network.
\newblock \emph{Journal of Hydrology}, 597:\penalty0 125776, 2021.
\newblock \doi{https://doi.org/10.1016/j.jhydrol.2020.125776}.

\bibitem[Wang and Xia(2015)]{Wang2015}
T.~Wang and Y.~Xia.
\newblock {W}hittle likelihood estimation of nonlinear autoregressive models with moving average residuals.
\newblock \emph{Journal of the American Statistical Association}, 110\penalty0 (511):\penalty0 1083--1099, 2015.
\newblock \doi{https://doi.org/10.1080/01621459.2014.946513}.

\bibitem[Wei et~al.(2022)Wei, Zhang, Jiang, and Huang]{WEI2022}
H.~Wei, H.~Zhang, H.~Jiang, and L.~Huang.
\newblock On the semi-varying coefficient dynamic panel data model with autocorrelated errors.
\newblock \emph{Computational Statistics \& Data Analysis}, 173:\penalty0 107458, 2022.
\newblock \doi{https://doi.org/10.1016/j.csda.2022.107458}.

\bibitem[Whittle(1953)]{whittle1953estimation}
P.~Whittle.
\newblock Estimation and information in stationary time series.
\newblock \emph{Arkiv f{\"o}r Matematik}, 2\penalty0 (5):\penalty0 423--434, 1953.
\newblock \doi{https://doi.org/10.1007/BF02590998}.

\bibitem[Wood(2017)]{wood2017generalized}
S.~N. Wood.
\newblock \emph{Generalized Additive Models: an Introduction with {R}}.
\newblock Chapman and Hall/CRC, 2017.
\newblock \doi{https://doi.org/10.1201/9781315370279}.

\bibitem[Yu et~al.(2017)Yu, Liu, Ma, and Bi]{yu2017pollution}
W.~Yu, Y.~Liu, Z.~Ma, and J.~Bi.
\newblock Improving satellite-based {PM2.5} estimates in {China} using {Gaussian} processes modeling in a {Bayesian} hierarchical setting.
\newblock \emph{Scientific Reports}, 7\penalty0 (1):\penalty0 7048, 2017.
\newblock \doi{https://doi.org/10.1038/s41598-017-07478-0}.

\bibitem[Zhang and Thorburn(2022)]{ZhangYifan2022Hmdi}
Y.~Zhang and P.~J. Thorburn.
\newblock Handling missing data in near real-time environmental monitoring: A system and a review of selected methods.
\newblock \emph{Future Generation Computer Systems}, 128:\penalty0 63--72, 2022.
\newblock \doi{https://doi.org/10.1016/j.future.2021.09.033}.

\bibitem[Zhou and Li(2011)]{ZHOU2011}
Y.~Zhou and W.~Li.
\newblock A review of regional groundwater flow modeling.
\newblock \emph{Geoscience Frontiers}, 2\penalty0 (2):\penalty0 205--214, 2011.
\newblock \doi{https://doi.org/10.1016/j.gsf.2011.03.003}.

\bibitem[Zhu and Zinde-Walsh(2009)]{Zhu2009AEP}
D.~Zhu and V.~Zinde-Walsh.
\newblock Properties and estimation of asymmetric exponential power distribution.
\newblock \emph{Journal of Econometrics}, 148\penalty0 (1):\penalty0 86--99, 2009.
\newblock \doi{https://doi.org/10.1016/j.jeconom.2008.09.038}.

\end{thebibliography}
\end{document}